\documentclass[journal,comsoc]{IEEEtran}

\usepackage{xspace}
\usepackage{amsmath}
\usepackage{float}
\usepackage[linesnumbered,ruled,vlined]{algorithm2e}
\usepackage[caption=false,font=footnotesize]{subfig}
\usepackage[justification=centering]{caption}
\usepackage{tabularx}
\usepackage{bbm}
\usepackage{xcolor}
\usepackage{booktabs}
\usepackage{multirow}
\usepackage{verbatim}
\usepackage{eurosym}
\usepackage{amsfonts}
\usepackage{paralist}

\ifCLASSINFOpdf
\usepackage[pdftex]{graphicx}
\else
\usepackage[dvips]{graphicx}
\fi
\graphicspath{{./images/}{./figs/}}
\DeclareGraphicsExtensions{.pdf,.jpg,.png}

\newtheorem{theorem}{Theorem}

\newcommand{\etal}{\textit{et al}.\xspace}

\newcommand{\ie}{\textit{i}.\textit{e}.\xspace}
\newcommand{\eg}{\textit{e}.\textit{g}.\xspace}
\newcommand{\sectionname}{Sec.}

\usepackage{dsfont}
\newcommand{\Prob}{\mathds{P}}

\newcommand{\heu}{MaxSR\xspace}
\newcommand{\bestfit}{Best-Fit\xspace}
\newcommand{\ratio}{cost/traffic ratio\xspace}

\newcommand{\Fig}[1]{\figurename~\ref{fig:#1}}

\newcommand{\Eq}[1]{\eqref{eq:#1}}
\newcommand{\Alg}[1]{Alg.~\ref{alg:#1}}
\newcommand{\Tab}[1]{Table~\ref{table:#1}}
\newcommand{\Line}[1]{Line~\ref{line:#1}}
\newcommand{\mLine}[2]{Lines~\ref{line:#1}-\ref{line:#2}}

\newcommand{\tvalue}[1]{\texttt{#1}}

\SetKwInput{KwInput}{Input}                
\SetKwInput{KwOutput}{Output}              

\SetArgSty{myargfont}
\SetKwInOut{Initialization}{Initialization}
\SetKwComment{Comment}{$\triangleright$\ }{}

\SetCommentSty{mycommfont}

\begin{document}

\title{Dynamic VNF Placement, Resource Allocation and Traffic Routing in 5G}

\author{Morteza Golkarifard,
	Carla Fabiana Chiasserini,
	Francesco Malandrino,
	and Ali Movaghar
	\thanks{M.~Golkarifard and A.~Movaghar are with Sharif University of Technology, Iran. F.~Malandrino and C.~F.~Chiasserini are with CNR-IEIIT and CNIT, Italy. C.~F.~Chiasserini is with Politecnico di Torino, Italy. }
}%

\maketitle
\begin{abstract}
    5G networks are going to support a variety of vertical services, with a diverse set of key performance indicators (KPIs), by using enabling technologies such as software-defined networking and network function virtualization. It is the responsibility of the network operator to efficiently allocate the available resources to the service requests in such a way to honor KPI requirements, while accounting for the limited quantity of available resources and their cost. A critical challenge is that requests may be highly varying over time, requiring a solution that accounts for their dynamic generation and termination. With this motivation, we seek to make joint decisions for request admission, resource activation, VNF placement, resource allocation, and traffic routing. We do so by considering real-world aspects such as the setup times of virtual machines, with the goal of maximizing the mobile network operator profit. To this end, first, we formulate a one-shot optimization problem which can attain the optimum solution for small size problems given the complete knowledge of arrival and departure times of requests over the entire system lifespan. We then propose an efficient and practical heuristic solution that only requires this knowledge for the next time period and works for realistically-sized scenarios. Finally, we evaluate the performance of these solutions using real-world services and large-scale network topologies. Results demonstrate that our heuristic solution performs better than state-of-the-art online algorithms and close to the optimum.
\end{abstract}

\section{Introduction}\label{sec:introduction}
5G networks are envisioned to support a variety of services belonging to vertical industries (\eg, autonomous driving, media, and entertainment) with a diverse set of requirements. 
Services are defined as a directed graph of virtual network functions (VNFs) with specific and varying key performance indicators (KPIs), \eg, throughput, and delay. 
Requests for these services arrive over time and mobile network operators (MNOs) are responsible for efficiently satisfy such a demand, by fulfilling their associated KPI while minimizing the cost for themselves.

As a result of the softwarization of 5G-and-beyond networks, enabled by software-defined networking (SDN) and network function virtualization (NFV), it is now feasible to use general-purpose resources (\eg, virtual machines) to implement the VNFs required by the different service. The decision on which resources to associate with which VNF and service is made by a network component called \textit{orchestrator}, as standardized by ETSI~\cite{etsimano}. Without loss of generalityi, we focus only on computational and communication resources (\eg, virtual machines and the links connecting them); notice, however, that our proposed framework is applicable to other resource types (\eg, storage).

The network orchestrator makes the following decisions~\cite{etsimano}: 
\begin{itemize}
	\item \textit{admission} of requests;
	\item \textit{activation/deactivation} of VMs;
	\item \textit{placement} of VNF instances therein;
	\item \textit{assignment} of CPU to VMs for running the hosted VNF instances;
	\item \textit{routing} of traffic through physical links.
\end{itemize}
These decisions are clearly mutually dependent, and therefore should be made {\em jointly}, in order to account for the -- often nontrivial -- ways in which they influence one another.
The focus of this paper is thus to consider the \textit{joint} requests admission, VM activation/deactivation, VNF placement, CPU assignment, and traffic routing problem in order to 
maximize the MNO profit, while considering:
\begin{itemize}
	\item the properties of each VNF,
	\item the KPI requirements of each service,
	\item the capabilities of VMs and PoPs (points of presence, \eg, datacenters) and their latency,
	\item the capacity and latency of physical links,
	\item the VMs setup times,
	\item the arrival and departure times of service requests.
\end{itemize}

As better discussed in \sectionname~\ref{sec:related-work}, some of these factors are simplified, or even neglected, in existing works on 5G orchestration. Notably, we account for the VM setup time, which becomes a significant factor in (for example) IoT applications, when requests are often short-lived. Ignoring setup (and tear-down) times can reduce the optimality of existing solutions. 

Furthermore, we account for the fact that different VNFs may have different levels of complexity, therefore, different quantities of computational resources may be needed to attain the same KPI target. Inspired by several works in the literature~\cite{agarwal2019vnf}, we model individual VNFs as queues and services as queuing networks. Critically, unlike traditional queuing networks, the quantity of traffic (\ie, the number of clients in queues) {\em can change} across queues, as VNFs can drop some packets (\eg, firewalls) or change the quantity thereof (\eg, video transcoders). Our model accounts for this important aspect by replacing traditional flow conservation constraints with a {\em generalized flow conservation} law, allowing us to describe arbitrary services with arbitrary VNF graphs.

Given this model, we formulate a one-shot optimization problem which, assuming perfect knowledge of future requests, allows us to maximize the MNO profit. Given the NP-hardness of such a problem and the fact that knowledge of future requests is usually not available, we propose \heu, an efficient heuristic algorithm which will be invoked periodically based on the knowledge of requests within each time period. The proposed method can achieve a near-optimal solution for large-scale network scenarios. We evaluate \heu compared to the optimum and other benchmarks using real-world services and different network scenarios.

In summary, the main contributions of this paper are as follows:
\begin{itemize}
	\item we propose a complete model for the main components of 5G, both in terms of vertical services (dynamic requests, VNFs, and services KPIs) and in terms of resources (\eg VMs and links); 
	\item our model accounts for the time variations of service requests, and dynamically allocates the computational and network resources while considering VMs setup times. It can also 
	accommodate a diverse set of VNFs in terms of computational complexity and KPI requirements, multiple VNF instances, and arbitrary VNF graphs with several ingress and egress VNFs, rather than a simple chain or directed acyclic graph (DAG); 
	\item we formulate a one-shot optimization problem as a Mixed-Integer Programming (MIP) to make a joint decision on VM state, VNF placement, CPU assignment, and traffic routing based on the complete requests statistics over the entire system lifespan;
	\item we propose \heu, an efficient near-optimal heuristic algorithm to solve the aforementioned problem based on the knowledge of the near future for large scale network scenarios;
	\item finally, we compare \heu with optimum and the online approach \bestfit, through extensive experiments using synthetic services and requests, and different network scenarios.
\end{itemize}

The rest of the paper is organized as follows.
\sectionname~\ref{sec:related-work} reviews related works. \sectionname~\ref{sec:system-model-problem-formulation} describes the system model and problem formulation, while \sectionname~\ref{sec:solution-strategy} clarifies our solution strategy. Finally, \sectionname~\ref{sec:numerical-results} presents our numerical evaluation under different network scenarios, and \sectionname~\ref{sec:conclusion} concludes the paper.

\section{Related Work\label{sec:related-work}}

Several works have addressed VNF placement and traffic routing, as exemplified by the survey paper \cite{yi2018comprehensive}. In most of these works, the problem is formulated as a Mixed Integer Linear Program (MILP) with a different set of objectives and constraints. Such an approach can yield exact solutions, but merely works for small instances; therefore, heuristic algorithms that offer a near-optimal solution have also been presented.

In particular, a first body of works provides a one-time VNFs placement, given the incoming service requests. Since this method leaves already placed VNFs intact, it can lead to a sub-optimal solution when the traffic varies over time. Examples of such an approach can be found in \cite{cohen2015near, gu2016comm, mechtri2016scalable, pham2017traffic, tajiki2019joint, pei2019efficient}, which aim at minimizing a cost function, \eg, operational cost, QoS degradation cost, server utilization, or a combination of them, and assume that there are always enough resources to serve the incoming requests. 
Among them, Cohen \etal~\cite{cohen2015near} propose an approximation algorithm to place sets of VNFs in an optimal manner, while approximating to the constraints by a constant factor.
Pham \etal~\cite{pham2017traffic} introduce a distributed solution based on a Markov approximation technique to place chains of VNFs where the cost enfolds the delay cost, in addition to the cost of traffic and server. 
\cite{tajiki2019joint}, instead, addresses the same problem but aims at minimizing the energy consumption, given constraints on end-to-end latency for each flow and server utilization.
Pei \etal~\cite{pei2019efficient} propose an online heuristic for this problem, by which VNF instances are deployed and connected using the shortest path algorithm, in order to minimize the number VNF instances and satisfy their end-to-end delay constraint.

Another thread of works focuses on an efficient admission policy that maximizes the throughput or revenue of admitted requests \cite{sallam2019joint, tahmasbi2018vspace, zhou2019efficient, kuo2018deploy}. 
In particular, Sallam \etal~\cite{sallam2019joint} formulate joint VNF placement and resource allocation problem to maximize the number of fully served flows considering the budget and capacity constraints. They leverage the sub-modularity property for a relaxed version of the problem and propose two heuristics with a constant approximation ratio. 
\cite{tahmasbi2018vspace} studies the joint VNF placement and service chain embedding problem, so as to maximize the revenue from the admitted requests. 
A similar problem is tackled in \cite{kuo2018deploy} and \cite{zhou2019efficient} but for an online setting where the requests should be admitted and served upon their arrival. 
Zhou \etal~\cite{zhou2019efficient}, on the other hand, first formulate a one-shot optimization problem over the entire system lifespan and then leverage the primal-dual method to design an online solution with a theoretically proved upper bound on the competitive ratio.

A different approach is adopted in \cite{fei2018adaptive, tang2019dynamic, jia2018online, li2016network, eramo2017approach, liu2017dynamic, huang2019max} where VNF placement can be readjusted through VNF sharing and migration, to optimally fit time-varying service demands.
\cite{fei2018adaptive} and \cite{tang2019dynamic} propose algorithms that properly scale over-utilized or under-utilized VNF instances based on the estimation of future service demands. 
Jia \etal~\cite{jia2018online} propose an online algorithm with a bounded competitive ratio that dynamically deploys delay constrained service function chains across geo-distributed datacenters minimizing operational costs. 

Request admission control has instead been considered in \cite{li2016network, eramo2017approach, liu2017dynamic, huang2019max}. More in detail, 
Li \etal~\cite{li2016network} propose a proactive algorithm that dynamically provisions resources to admit as many requests as possible with a timing guarantee. 
Similarly, \cite{eramo2017approach} admits requests and places their VNFs in the peak interval, but minimizes the energy cost of VNF instances by migration and turning off empty ones in the off-peak interval. 
Liu \etal~\cite{liu2017dynamic} envision an algorithm that maximizes the service provider's profit by periodically admitting new requests and rearranging the current-served ones, while accounting for the operational overhead of migration. 
Finally, leveraging VNF migration and sharing, \cite{huang2019max} proposes an online algorithm to maximize throughput while minimizing service cost and meeting latency constraints.

Relevant to our work are also studies that target specifically 5G systems, although they merely consider the link delay and neglect processing delays in the servers.
An example can be found in \cite{agarwal2019vnf}, which models VMs as M/M/1 PS queues, and proposes a MILP and a heuristic solution to minimize the average service delay, while meeting the constraints on the links and host capacities. 
The works in \cite{malandrino2019optimization} and \cite{malandrino2019reduce} aim instead to minimize, respectively, the operational cost and the energy consumption of VMs and links while ensuring end-to-end delay KPI. \cite{malandrino2019reduce} also allows for VNF sharing and studies the impact of applying priorities to different services within a shared VNF.
Zhang \etal~\cite{zhang2019adaptive} tackle the request admission problem to maximize the total throughput, neglecting instead queuing delay at VMs.

We remark that most of the above works present proactive approaches, and only deal with either cost minimization or request admission. On the contrary, we focus on dynamic resource activation, VNF placement, and CPU assignment to maximize the revenue from admitted requests over the entire system lifespan, while minimizing the deployment costs and accounting for some practical issues. Our proactive MILP formulation of the problem extends existing models by accounting for the maximum end-to-end delay as the main KPI, while our heuristic is a practical and scalable solution, which periodically admits new requests and readjusts the existing VNF deployment. To the best of our knowledge, this is the first dynamic solution for service orchestration in 5G networks.

\section{System Model and Problem Formulation}
\label{sec:system-model-problem-formulation}
In this section, first we describe our system model supported by a simple example. Later, we formulate the \textit{joint} requests admission, VM activation, VNF placement, CPU assignment, and traffic routing problem; a discussion of the problem time complexity follows. The frequently used notation is summarized in \Tab{notation}.

\begin{table*}[t]
	\caption{Notation (sets, variables, and parameters)}
	\label{table:notation}
	\noindent
	\begin{tabular}[t]{@{}p{0.5\textwidth}@{}p{0.5\textwidth}@{}}
		\begin{tabularx}{0.49\textwidth}[t]{@{}lX@{}}
			\toprule
			Symbol & Description \\ 
			\midrule
			$\mathcal{D}$ & Set of datacenters\\
			$\mathcal{E}$ & Set of physical links\\
			$\mathcal{K}$ & Set of service requests\\
			$ \mathcal{L}$ & Set of logical links \\
			$\mathcal{M}$ & Set of VMs \\
			$\mathcal{P}$ & Set of end-to-end paths \\
			$\mathcal{Q}$ & Set of VNFs\\
			$\mathcal{S}$ & Set of services\\
			$\mathcal{T}$ & Set of time steps \\
			$\mathcal{W}_s$ & Set of paths from ingress VNFs to egress VNFs in VNF graph of service $s$ \\
			\midrule
			$A(k,m,q,t)$& Whether to deploy VNF $q$ of service request $k$ at VM $m$ at time $t$\\
			$D(k,m,q,t)$& Traffic departing VM $m$ for VNF $q$ of service request $k$ at time $t$\\
			$F(k,l,q_1,q_2,t)$& Equal to $1$ when $\rho(k,l,q_1,q_2,t)>0$ \\
			$I(k,m,q,t)$& Traffic entering VM $m$ for VNF $q$ of service request $k$ at time $t$\\
			$L(e,t)$& Traffic on physical link $e$ at time $t$\\
			$O(m,t)$& Whether VM $m$ is \textit{active} at time $t$ \\
			$R(m,t)$& Average time for a request to be processed at VM $m$ at time $t$\\
			$U(m,t)$& Whether VM $m$ is \textit{turning-on} at time $t$ \\
			$V(k,t)$& Whether service request $k$ is \textit{active} at time $t$ \\
			$\mu(k,m,q,t)$& Service rate to assign to VM $m$ for VNF $q$ of service request $k$ at time $t$\\
			$\rho(k,l,q_1,q_2,t)$& Fraction of traffic from VNF $q_1$ to $q_2$ of service request $k$, through logical link $l$ at time $t$ \\
		\end{tabularx}
		&
		\begin{tabularx}{0.49\textwidth}[t]{@{}lX}
			\toprule
			Symbol & Description \\ \midrule
			$B(e)$& Bandwidth of physical link $e$\\
			$C_{\textit{dc}}(d)$& Computational capacity of datacenter $d$\\
			$C_{\textit{vm}}(m)$& Computational capacity of VM $m$\\
			$D_{\textit{QoS}}(s)$& Target delay for service $s$\\
			$D_{\textit{log}}(l)$& Delay of logical link $l$\\
			$D_{\textit{phy}}(e)$& Delay of physical link $e$\\
			$N(s,q)$& Maximum number of instances for VNF $q$ of service $s$\\
			$X_{\textit{cpu}}(m)$& Cost for VM $m$ to process one unit of computation in one time step \\
			$X_{\textit{idle}}(m)$& Fixed cost incurred when VM $m$ is \textit{turning-on} or \textit{active} in one time step \\
			$X_{\textit{link}}(e)$& Cost of data transmission through physical link $e$ in one time step \\
			$X_{\textit{rev}}(s)$& Revenue from serving one traffic unit of service $s$ \\
			$\Lambda(s,q_1,q_2)$& Traffic from VNF $q_1$ to $q_2$ for service $s$ \\
			$\Prob(s, q_1, q_2)$& Probability that traffic processed at VNF $q_1$ is forwarded to VNF $q_2$ of service $s$\\
			$\alpha(s, q)$& Ratio of outgoing traffic to incoming traffic for VNF $q$ of service $s$\\
			$\lambda_{\textit{new}}(s)$& New traffic for service $s$ \\
			$\omega(q)$ & Computation capability required for one traffic unit at VNF $q$\\
			$t_{\textit{arv}}(k)$& Arrival time of service request $k$ \\
			$t_{\textit{dpr}}(k)$& Departure time of service request $k$\\
		\end{tabularx}
		\\ \bottomrule
	\end{tabular}
\end{table*}

\subsection{System Model}
\label{sec:system-model}

\textbf{Physical infrastructure.}
Let $\mathcal{G}=(\mathcal{M}, \mathcal{E})$ be a directed graph representing the physical infrastructure network, where each node $m \in \mathcal{M}$ is either a VM or a network node (\ie, a router or a switch). A VM $m$ has maximum computational capacity $C_{\textit{vm}}(m)$. Set $\mathcal{E}$ denotes the physical links connecting the network nodes. We define $B(e)$ and $D_{\textit{phy}}(e)$ as, respectively, the bandwidth and delay of physical link $e \in \mathcal{E}$. Time is discretized into {\em steps}, $\mathcal{T}=\{1,2,…,T\}$, and we assume that at every time step a VM 
may be in one of the following states: \textit{terminated}, \textit{turning-on}, or \textit{active}. Specifically, VMs can only be used when they are active, and they need to be turned-on one time step before being active. 
Based on the measurements reported in~\cite{tang2019dynamic}, we also consider the traffic flow migration time to be negligible with respect to the VM setup time.

Each VM can host one VNF and belongs to a datacenter $d \in \mathcal{D}$; 
we denote the available amount of computational resources in datacenter $d$ by $C_{\textit{dc}}(d)$ and the 
set of VMs within $d$ with $\mathcal{M}_{d}$. 
In the physical graph $G$, physical links within datacenters are assumed to be ideal, \ie, they have no capacity limit and zero delay. Let logical link $l \in \mathcal{L}$ be a sequence of physical links connecting two VMs, $\textit{src}(l)$ and destination $\textit{dst}(l)$, 
then we define end-to-end path $p \in \mathcal{P}$ as a sequence of logical links. 

\textbf{Services.} 
We represent each service $s \in \mathcal{S}$ with a \textit{VNF Forwarding Graph (VNFFG)}, where the nodes are VNFs $q \in \mathcal{Q}$, and the directed edges show how traffic traverses the VNFs.
VNFFG can be any general graph with possibly several ingress and egress VNFs. 
We denote the total new traffic, entering the ingress VNFs of service $s$, by $\lambda_{\textit{new}}(s)$.
A traffic packet of service $s$, processed in VNF $q_1$, is forwarded to VNF $q_2$ with probability of $\Prob(s,q_1,q_2)$. Similarly, $\Prob(s,\circ,q)$ is the probability that a new traffic packet of service $s$ starts getting service in ingress VNF $q$, and $\Prob(s,q,\circ)$ is the probability that a traffic packet of service $s$, already served at egress VNF $q$, departs service $s$. 
For each service $s$, we consider its target delay, $D_{\textit{QoS}}(s)$, as the most critical KPI, specifying the maximum tolerable end-to-end delay for the traffic packets of $s$.

VNFs can have different processing requirements depending on their computational complexity. We denote by $\omega(q)$ the computational capability that VNF $q$ needs to process one unit of traffic. 
Some VNFs may not find sufficient resources on a single VM to completely serve the traffic while satisfying the target delay. 
Thus, multiple instances can be created, with $N(s,q)$ being the maximum number of instances of VNF $q$ at each point in time. Instances of the same VNF can be deployed either within the same datacenter or at different datacenters; in the latter case, the traffic between each pair of VNFs must be splitted through different logical links that connect the VMs running the corresponding VNF instances.

Different requests for the same services may arrive over time; we denote with $K_s$ the set of all service requests for service $s$, and characterize the generic service request $k \in \mathcal{K}$ with its arrival time $t_{\textit{arv}}(k)$ and departure time $t_{\textit{dpr}}(k)$. 
Due to slice {\em isolation} requirements~\cite{alliance2016description}, we assume that the VNF instances of different service requests are not shared with other service requests. 

\begin{figure}[tbh]
	\centering
	\subfloat[\label{fig:model-exm1}]{{\includegraphics[width=0.7\linewidth]{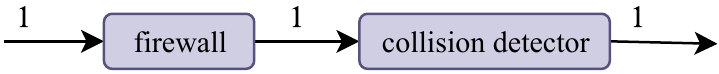} }}
	\;
	\subfloat[\label{fig:model-exm2}]{{\includegraphics[width=0.8\linewidth]{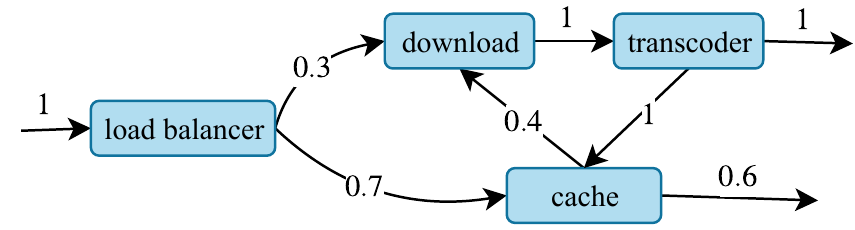} }}
	\;
	\subfloat[\label{fig:model-exm3}]{{\includegraphics[width=0.9\linewidth]{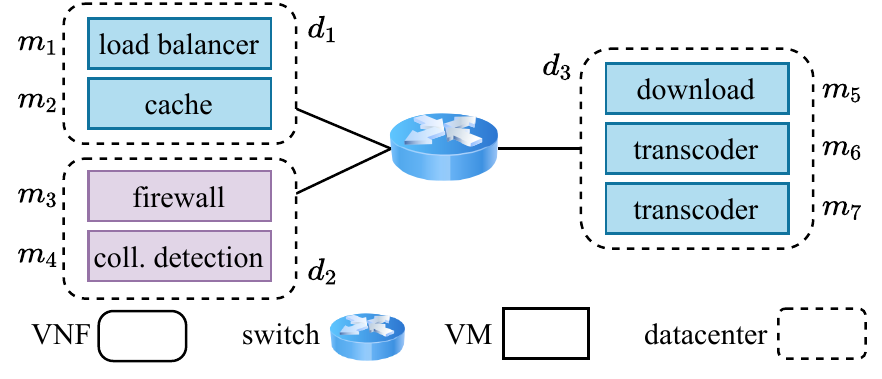} }}
	\caption{VNFFG of \protect\subref{fig:model-exm1} vehicle collision detection (VCD) service and \protect\subref{fig:model-exm2} video on-demand (VoD) service. The number on edges represents transition probability of traffic packets. \protect\subref{fig:model-exm3} Physical graph including three datacenters connected using a switch.}
	\label{fig:model-exm}
\end{figure}

\textbf{Example.}
\Fig{model-exm} represents a possible deployment of two sample services, vehicle collision detection (VCD) and video on-demand (VoD), on the physical graph (\Fig{model-exm3}) in a single time step. VCD is a low-latency service with a very low target delay $D_{\textit{QoS}}$, and VoD is a traffic intensive service with a high $\lambda_{\textit{new}}$. \Fig{model-exm1} and \Fig{model-exm2} depict the VNFFGs of the VCD and VoD services, respectively, where the numbers on the edges represent the transition probability of traffic packets between corresponding VNFs. The physical graph contains a set of datacenters $\mathcal{D}=\{ d_1, d_2, d_3 \}$ with computational capability $C_{\textit{dc}}$. Datacenters are connected to each other using a switch and physical links with bandwidth $B$ and a latency $D_{\textit{phy}}$. 
VMs within each datacenter are denoted by sets $\mathcal{M}_{d_1}=\{ m_1, m_2 \}$, $\mathcal{M}_{d_2}=\{ m_3, m_4 \}$, and $\mathcal{M}_{d_3}=\{ m_5, m_6, m_7 \}$, each with computational capability $C_{\textit{vm}}$. 
As depicted in~\Fig{model-exm3}, service VCD is deployed within datacenter $d_2$ to avoid inter-datacenter network latency. Service VoD is deployed across datacenter $d_1$ and third-party datacenter $d_3$. VNF \textit{transcoder}, having high computational complexity $\omega$, requires two instances in datacenters $d_3$ to fully serve the traffic.

\subsection{Problem Formulation}
\label{sec:problem-formulation}

In this section, we first describe the decisions that have to be made to map the service requests onto network resources. Then we formalize the system constraints and the objective using the model presented in \sectionname~\ref{sec:system-model}, along with the decision variables we define.
In general, given the knowledge of the future arrival and departure times of service requests, we should make the following decisions:
\begin{itemize}
	\item service request activation, \ie, when service requests get served; 
	\item VM activation/deactivation, \ie, when VMs are set up or terminated;
	\item VNF instance placement, \ie, which VMs have to run VNF instances;
	\item CPU assignment, \ie, how much computational capability shall be assigned to a VM to run the deployed VNF;
	\item traffic routing, \ie, how traffic between VNFs is routed through physical links.
\end{itemize}

\textbf{Service request activation.}
Let binary variable $V(k,t) \in \{0,1\}$ denote whether service request $k$ is being served at time $t$. Once admitted, a service request has to be provided for all its lifetime duration. Given service request arrival time $t_{\textit{arv}}(k)$ and departure time $t_{\textit{dpr}}(k)$, this translates into: 
\begin{multline}
V(k,t)=0, \hfill \forall k \in \mathcal{K}, t \in \mathcal{T}: t < t_{\textit{arv}}(k) \vee t \ge t_{\textit{dpr}}(k). \label{eq:dynamic1}
\end{multline}

\textbf{VNF instances.} 
The following constraint limits the number of deployed instances of VNF $q$ of any service request $k \in \mathcal{K}_s$ to be less than $N(s,q)$ at any point in time:
\begin{multline}
\sum_{m \in \mathcal{M}}{A(k,m,q,t)} \le N(s,q),
\\ \forall t \in \mathcal{T}, s \in \mathcal{S}, k \in \mathcal{K}_s, q \in \mathcal{Q}, \label{eq:vnf-instance} 
\end{multline}
where binary variable $A(k,m,q,t)$ represents whether VNF $q$ of service request $k$ is placed on VM $m$ at time $t$. 
The network slice isolation property of 5G networks prevents VNF sharing among requests for different services. In addition, at most one VNF instance can be deployed on any VM, \ie, 
\begin{multline}
\sum_{k \in \mathcal{K}}{\sum_{q \in \mathcal{Q}}{A(k,m,q,t)}} \le 1,\hfill \forall m \in \mathcal{M}, t \in \mathcal{T}. \label{eq:vnf-sharing}
\end{multline}

\textbf{VM states.} 
We define two binary variables $U(m,t)$ and $O(m,t)$ to represent whether VM $m$ is \textit{turning-on} or \textit{active} at time $t$, respectively. We formulate a simple constraint to prevent VMs from being concurrently turning-on and active at any time, \ie,
\begin{multline}
O(m,t) + U(m,t) \le 1, \hfill \forall m \in \mathcal{M}, t \in \mathcal{T}. \label{eq:vm_state}
\end{multline}
The following constraint enforces that VM $m$ can be \textit{active} at time $t$ only if it 
has been turning-on or active in the previous time step:
\begin{multline}
O(m,t) \le O(m,t-1) + U(m,t-1), \hfill \forall m \in \mathcal{M}, t \in \mathcal{T}. \label{eq:vm-setup} 
\end{multline}
VMs are able to run VNFs only when they are active, \ie, 
\begin{multline}
\sum_{k \in \mathcal{K}}{\sum_{q \in \mathcal{Q}}{A(k,m,q,t)}} \le O(m,t),\hfill \forall m \in \mathcal{M}, t \in \mathcal{T}. \label{eq:vm-availability} 
\end{multline}

\textbf{Computational capacity.}
Let real variable $\mu(k,m,q,t)$ represent the service rate assigned to VM $m$ to run VNF $q$ of service request $k$ at time $t$. Multiplying it by $\omega(q)$, we have the amount of computation capability assigned to VM $m$ to run VNF $q$ at time $t$. 
The limited computational capability of datacenters and VMs denoted, respectively, by $C_{\textit{dc}}(d)$ and $C_{\textit{vm}}(m)$, should not be exceeded at any point in time. We describe such a limitation by imposing:
\begin{multline}
\sum_{m \in \mathcal{M}_d}{\sum_{k \in \mathcal{K}}{\sum_{q \in \mathcal{Q}}{
			\mu(k,m,q,t) \cdot \omega(q)}}} \le C_{\textit{dc}}(d),
\\ \forall t \in \mathcal{T}, d \in \mathcal{D}, \label{eq:dc-capacity}
\end{multline}
where the sum on the left-hand side of the inequality is over all VMs within datacenter $d$.
Similarly, for the VMs we have
\begin{multline}
\mu(k,m,q,t) \cdot \omega(q) \le A(k,m,q,t) \cdot C_{\textit{vm}}(m),
\\ \forall t \in \mathcal{T}, k \in \mathcal{K}, q \in \mathcal{Q}, m \in \mathcal{M}, \label{eq:vm-capacity}
\end{multline}
where $A(k,m,q,t)$ on the right-hand side of the inequality enforces zero service rate for VM $m$ when no VNF is placed therein. 

\textbf{KPI target fulfillment.}
Whenever a service request is being served, \ie, $V(k,t)=1$, all the traffic in the corresponding VNFFG should be carried by the underlying physical links. The following constraint ensures this condition for the traffic between each pair of VNFs at any point in time: 
\begin{multline}
\sum_{l \in \mathcal{L}}{\rho(k,l,q_1,q_2,t)} \ge V(k,t), \\
\forall t \in \mathcal{T}, s \in \mathcal{S}, k \in \mathcal{K}_s, q_1, q_2 \in \mathcal{Q} : \Prob(s, q_1, q_2)>0. \label{eq:splitting-intermediate} 
\end{multline}
Real variable $\rho(k,l,q_1,q_2,t)$ shows the fraction of traffic from VNF $q_1$ to $q_2$ of service request $k$ that is routed through logical link $l$ at time $t$. As mentioned, the traffic flow from VNF $q_1$ to VNF $q_2$ may be splitted into several logical links (see Eq. \Eq{vnf-instance}). Moreover, since we consider multi-path routing, there may be multiple logical links between each pair of VNF instances. Therefore, constraint \Eq{splitting-intermediate} implies that for any service request $k$ requesting traffic from VNF $q_1$ to $q_2$ (\ie, $\Prob(s, q_1, q_2)>0$), the sum of all fractional traffic going though any logical link, should be equal to $1$ at any time when the service request is being served. 

The above constraint does not include ingress and egress traffic. To account for such contributions, we need to introduce \textit{dummy} nodes in the VNFFG and the physical graph. We add an end-point \textit{dummy} VNF, $\circ$ in every VNFFG, which is directly connected to all ingress and egress VNFs and a \textit{dummy} VM in the physical graph which is directly connected to all VMs. We define $\mathcal{L}_{\circ}$ as the set of \textit{dummy} logical links which start from or end at the \textit{dummy} VM. We assume that \textit{dummy} logical links are ideal, \ie, they have no capacity limit and zero delay and cost. We can now formulate the associated traffic constraints as:
\begin{multline}
\sum_{l \in \mathcal{L}_{\circ}}{\rho(k,l,\circ,q,t)} \ge V(k,t), \\ \forall t \in \mathcal{T}, s \in \mathcal{S}, k \in \mathcal{K}_s, q \in \mathcal{Q}
: \Prob(s, \circ, q)>0, \label{eq:splitting_ingress} 
\end{multline}
\begin{multline}
\sum_{l \in \mathcal{L}_{\circ}}{\rho(k,l,q,\circ,t)} \ge V(k,t), \\
\forall t \in \mathcal{T}, s \in \mathcal{S}, k \in \mathcal{K}_s, q \in \mathcal{Q} : \Prob(s, q, \circ)>0, \label{eq:splitting_egress}
\end{multline}
where $\rho(k,l,\circ,q,t)$ and $\rho(k,l,q,\circ,t)$ are the fraction of new traffic entering ingress VNF $q$ and the fraction of traffic departing from egress VNF $q$, respectively, going through logical link $l$ at time $t$.

\textbf{Placement.} We can now 
correlate the routing decisions $\rho$ and the placement decisions $A$ as
\begin{multline}
\rho(k,l,q_1,q_2,t) \le A(k,m,q_2,t),
\\ \forall t \in \mathcal{T}, k \in \mathcal{K}, q_1 \in \mathcal{Q} \cup \{ \circ \}, q_2 \in \mathcal{Q}, \\
m \in \mathcal{M}, l \in \mathcal{L} \cup \mathcal{L}_{\circ} : \textit{dst}(l)=m. \label{eq:placement-in} 
\end{multline}
The above constraint implies that whenever there is an incoming traffic to VNF $q_2$ through logical link $l$ whose destination is VM $m$, \ie, $\textit{dst}(l)=m$, VNF $q_2$ is deployed at VM $m$. Similarly, whenever there is an outgoing traffic from VNF $q_1$ through logical link $l$ whose source is VM $m$, \ie, $\textit{src}(l)=m$, VNF $q_1$ is deployed at VM $m$:
\begin{multline}
\rho(k,l,q_1,q_2,t) \le A(k,m,q_1,t),
\\ \forall t \in \mathcal{T}, k \in \mathcal{K}, q_1 \in \mathcal{Q}, q_2 \in \mathcal{Q} \cup \{ \circ \} \\
m \in \mathcal{M}, l \in \mathcal{L} \cup \mathcal{L}_{\circ} : \textit{src}(l)=m. \label{eq:placement-out} 
\end{multline}

\textbf{System stability.} 
Let $\lambda(s,q)$ denote the total incoming traffic of VNF $q$ of service $s$. $\lambda(s,q)$ equals the sum of ingress traffic and the traffic coming from other VNFs to VNF $q$ of service $s$: 
\begin{multline}
\lambda(s,q) = \lambda_{\textit{new}}(s) \cdot \Prob(s,\circ,q) + \\ + \sum_{q_1 \in \mathcal{Q} \backslash \{q\}}{\lambda(s,q') \cdot \Prob(s,q',q)}. \label{eq:lambda}
\end{multline}
Using $\lambda(s,q)$, the amount of traffic from VNF $q_1$ to VNF $q_2$ of service $s$ can be represented as:
\begin{equation}
\Lambda(s,q_1,q_2) = \lambda(s,q_1) \cdot \Prob(s,q_1,q_2). \hfill \label{eq:Lambda}
\end{equation}
We can now define an auxiliary variable to represent the incoming traffic of VNF $q$ of service request $k$, which enters VM $m$ at time $t$:
\begin{multline}
I(k,m,q,t) = \sum_{\substack{q' \in \mathcal{Q} \cup \{ \circ \}}}{
	\sum_{\substack{l \in \mathcal{L} \cup \mathcal{L}_{\circ}:\\ \textit{dst}(l)=m}}{
		\rho(k,l,q',q,t) \cdot \Lambda(s,q', q)}},
\\ t \in \mathcal{T}, s \in \mathcal{S}, k \in \mathcal{K}_s, q \in \mathcal{Q}, m \in \mathcal{M}, \label{eq:incoming-traffic}
\end{multline}
where the summation is over all logical links ending at VM $m$. Finally, we describe the \textit{system stability} requirement, which imposes the incoming traffic not to exceed the assigned service rate for each VNF $q$ of service request $k$ on VM $m$, at any point in time:
\begin{multline}
I(k,m,q,t) \le \mu(k, m, q, t),
\\ \forall t \in \mathcal{T}, k \in \mathcal{K}, q \in \mathcal{Q}, m \in \mathcal{M}. \label{eq:stability} 
\end{multline}

\textbf{Generalized flow conservation.}
Our model captures the possibility of having VNFs for which, due to processing, the amount of incoming and that of outgoing traffic are different. We define the scaling factor $\alpha(s, q)$ as the ratio of outgoing traffic to incoming traffic for VNF $q$ of service $s$:
\begin{multline}
\alpha(s,q) = \frac{\sum_{q' \in \mathcal{Q} \cup \{ \circ \}}{\Lambda(s,q,q')}}{\sum_{q' \in \mathcal{Q} \cup \{ \circ \}}{\Lambda(s,q',q)}}, 
\hfill s \in \mathcal{S}, q \in \mathcal{Q}. \label{eq:alpha}
\end{multline}
We also define auxiliary variable $D(k,m,q,t)$ to represent the outgoing traffic of VNF $q$ of service request $k$ departing VM $m$ at time $t$:
\begin{multline}
D(k,m,q,t) = \sum_{\substack{q' \in \mathcal{Q} \cup \{ \circ \}}}{
	\sum_{\substack{l \in \mathcal{L} \cup \mathcal{L}_{\circ}:\\ \textit{src}(l)=m}}{
		\rho(k,l,q,q',t) \cdot \Lambda(s,q, q')}},
\\t \in \mathcal{T}, s \in \mathcal{S}, k \in \mathcal{K}_s, q \in \mathcal{Q}, m \in \mathcal{M}, \label{eq:outgoing-traffic}
\end{multline}
where the right-hand side enfolds all traffic flowing through logical links starting from VM $m$. 
We can then formulate the \textit{generalized flow conservation law} for each VNF $q$ of service request $k$ on VM $m$ at time $t$:
\begin{multline}
D(k,m,q,t) = \alpha(s, q) \cdot I(k,m,q,t),\\ \forall t \in \mathcal{T}, s \in \mathcal{S}, k \in \mathcal{K}_s, q \in \mathcal{Q}, m \in \mathcal{M}, \label{eq:flow-conservation}
\end{multline}
which implies that for each VNF $q$ of service request $k$ on VM $m$, at any time, the outgoing traffic is equal to the incoming traffic multiplied by the scaling factor $\alpha(s, q)$. 

\textbf{Latency.} 
End-to-end network latency for a traffic packet of a service request is the time it takes to the packet to be served by all VNFs along the path from the ingress to the egress VNFs. Such a latency includes two contributions, namely, the network delay between pairs of VMs on which subsequent VNFs are deployed and the processing time at the VNFs themselves. The former can be defined based on the delay of the logical links $l$, denoted by $D_{\textit{log}}(l)$. Such a delay is the sum of the delay of the underlying physical links: 
\begin{equation}
D_{\textit{log}}(l) = \sum_{e \in l}{D_{\textit{phy}}(e)}. \hfill
\end{equation}
We also introduce binary variable $F(k,l,q_1,q_2,t)$ to represent whether logical link $l$ is used for routing the traffic from VNF $q_1$ to $q_2$ of service request $k$ at time $t$. $F$ can be described as
\begin{multline}
\rho(k,l,q_1,q_2,t) \le F(k,l,q_1,q_2,t), 
\\ \forall t \in \mathcal{T}, k \in \mathcal{K}, q_1,q_2 \in \mathcal{Q}, l \in \mathcal{L}. \label{eq:ensure_F} 
\end{multline}
The traffic packets in the VNFFG follow a path $p$ of logical links in the underlying physical graph, which connect all VNFs in the VNFFG. Let $w \in \mathcal{W}_s$ be the sequence of VNFs, from an ingress VNF to an egress VNF in the VNFFG of service $s$. The network delay of traffic packets of service request $k$, which traverse the VNFs as specified by $w$ and go through the links belonging to $p$, is given by:
\begin{equation}
\sum_{(q_1,q_2) \in w}{
	\sum_{l \in p}{
		F(k,l,q_1,q_2,t) \cdot D_{\textit{log}}(l)
	}
}. \hfill
\end{equation}

The processing time of VM $m$, denoted by $R(m,t)$, is the time it takes for a traffic packet to be completely processed in the VM. Modeling each VM as a queue with discipline PS (or, equivalently, FIFO), the processing time of VM $m$ at time $t$ is~\cite{agarwal2019vnf}:
\begin{multline}
R(m,t) = \frac{1}{\sum_{k \in \mathcal{K}}{\sum_{q \in \mathcal{Q}}{\left( \mu(k,m,q,t) - I(k,m,q,t) \right)}}},	
\\ m \in \mathcal{M},t \in \mathcal{T}. \label{eq:processing_time}
\end{multline}
Then, the processing time incurred by the traffic packets following the VNF sequence $w$, is given by: 
\begin{equation}
\sum_{q \in w}{
	\sum_{m \in p}{
		A(k,m,q,t) \cdot R(m,t)
	}
}. \hfill \label{eq:processing_delay}
\end{equation}
Finally, the experience delay must be less than the target delay, \ie,
\begin{multline}
\sum_{(q_1,q_2) \in w}{
	\sum_{l \in p}{
		F(k,l,q_1,q_2,t) \cdot D_{\textit{log}}(l)
	}
}
+ \\ +
\sum_{q \in w}{
	\sum_{m \in p}{
		A(k,m,q,t) \cdot R(m,t)
	}
}
\le D_{\textit{QoS}}(s), \\
\forall t \in \mathcal{T}, s \in \mathcal{S}, k \in \mathcal{K}_s, w \in \mathcal{W}_s, p \in \mathcal{P}. \label{eq:latency} 
\end{multline}

\textbf{Link capacity.} 
The traffic on any physical link should not exceed the maximum link capacity, $B(e)$. To formalize this constraint, we define the auxiliary variable $L(e,t)$ to represent the traffic on physical link $e$ at time $t$. This variable is equal to the total traffic between each pair of VNFs which goes through the logical link $l$ containing the physical link $e$: 
\begin{multline}
L(e,t) = \sum_{s \in \mathcal{S}}{
	\sum_{k \in \mathcal{K}_s}{
		\sum_{\substack{q_1, q_2 \in \mathcal{Q}}}{\sum_{e \in l}{
				\Lambda(s,q_1,q_2) \cdot \rho(k,l,q_1,q_2,t) }}}}. \label{eq:link_load}
\end{multline}
The link capacity constraint is expressed as
\begin{multline}
L(e,t) \le B(e),\hfill \forall e \in \mathcal{E}, t \in \mathcal{T}. \label{eq:bandwidth}
\end{multline}

\textbf{Objective.} 
The goal of the optimization problem is to maximize the service revenue while minimizing the total cost. The revenue obtained by serving one unit of traffic of service~$s$ is indicated as $X_{\textit{rev}}(s)$; we assume such a quantity to be 
inversely proportional
to the target delay of service $s$, \ie, $1/D_{\textit{QoS}}(s)$. This implies that serving services with lower target delay yields higher revenue for the MNO. The total revenue is expressed as
\begin{equation}
R = \sum_{t \in \mathcal{T}}{
	\sum_{s \in \mathcal{S}}{
		\sum_{k \in \mathcal{K}_s}{
			X_{\textit{rev}}(s) \cdot V(k,t) \cdot \lambda_{\textit{new}}(s) }}}.
\label{eq:cost-rev}
\end{equation}
The total cost is the sum of the transmission cost in physical links, computational and idle costs in VMs, which are described, respectively, as:
\begin{align}
C_{\textit{link}} &= \sum_{t \in \mathcal{T}}{
	\sum_{e \in \mathcal{E}}{
		X_{\textit{link}}(e) \cdot L(e,t) }},
\label{eq:cost-link}
\\
C_{\textit{cpu}} &= \sum_{t \in \mathcal{T}}{
	\sum_{m \in \mathcal{M}}{\sum_{k \in \mathcal{K}}{\sum_{q \in \mathcal{Q}}{
				X_{\textit{cpu}}(m) \cdot \mu(k,m,q,t) \cdot \omega(q) }}}},
\label{eq:cost-cpu}
\\ 
C_{\textit{idle}} &= \sum_{t \in \mathcal{T}}{
	\sum_{m \in \mathcal{M}}{
		X_{\textit{idle}}(m) \cdot (U(m,t) + O(m,t)) }}.
\label{eq:cost-idle}
\end{align}
The above costs are expressed per unit of time and depend, respectively, on a proportional cost $X_{\textit{link}}(e)$ paid for each physical link $e$ per unit of traffic, a proportional cost $X_{\textit{cpu}}(m)$ for each VM $m$ paid per unit of computation, and a fixed cost $X_{\textit{idle}}(m)$ for each VM $m$ paid if VM $m$ is \textit{turning-on} or \textit{active}. Finally, we write our objective as:
\begin{equation}
\max \left[ R - (C_{\textit{link}} + C_{\textit{cpu}} + C_{\textit{idle}}) \right] \,.
\label{eq:obj}
\end{equation}

\subsection{Problem Complexity}
\label{sec:problem-complexity}
The problem of jointly making decisions about VM activation, VNF placement, CPU assignment, and traffic routing formulated above contains both integer and real decision variables, hence it is non convex. In the following, we prove that the problem is NP hard, through a reduction from the weight constrained shortest path problem (WCSPP) to a simpler version of our own.
\begin{theorem}
	The problem mentioned in \sectionname~\ref{sec:system-model} is NP-hard when the objective value is greater than zero.
\end{theorem}
\begin{IEEEproof}
	We reduce an NP-hard problem, called weight constrained shortest path problem (WCSPP)~\cite{dumitrescu2001algorithms}, to our problem. Given a graph $G(V,E)$, and the cost and weight associated with the edges, the WCSPP asks to find the minimum cost route between two specified nodes while ensuring that the total weight is less than a given value.
	We consider a special case of our problem where only one service request with a chain of two VNFs arrives at $t=1$ and departs in the next time step. We set the maximum number of instances for both VNFs to one. There are only two VMs in the physical infrastructure, with $C_{\textit{vm}}(m)=\infty$ and $X_{\textit{cpu}}(m) = X_{\textit{idle}}(m) = 0$; the remaining are network nodes. We set $C_{\textit{dc}}(d)=\infty, \forall d\in \mathcal{D}$. Then, it is easy to see that WCSPP is equivalent to the special case of our problem when the objective value is greater than zero.
\end{IEEEproof}
Beside complexity, solving the problem formulated in \sectionname~\ref{sec:problem-formulation} assumes that the entire knowledge of arrival and departure times of all service requests is available, which is not realistic in many scenarios. 
As detailed below, to cope with this issue, our strategy is to periodically solve our problem, with each problem instance leveraging only the information about the past and the current service requests.

\section{The \heu Solution\label{sec:solution-strategy}}

In light of the problem complexity discussed above, we propose a heuristic solution called \heu, which makes decisions (i) only concerning a subsequent time interval encompassing the present and the near future, which can be predicted with high accuracy~\cite{bega2020deepcog}, (ii) based on the knowledge of the service requests occurring within such time interval. 
More precisely, starting from time step $t$, \heu makes decisions concerning the current service requests and accounting for a time horizon $H$, \ie, extending till $t+H$. After $\tau$ time steps, where $\tau \le H$, \heu is executed again accounting for the next time interval, \ie, $\left[t+\tau, t+\tau+H\right)$.
Note that, although decisions are made accounting for a time horizon equal to $H$, they will be enacted just until the next execution of \heu, \ie, they hold, in practice, only for $\tau$. 
Even with such a limited time horizon, directly solving the problem defined in \sectionname~\ref{sec:problem-formulation} is still NP hard. To walk around this limitation, at every execution, \heu processes the service requests received in the last $\tau$ time steps sequentially, \ie, one request at a time. 
In the following, we provide an overview of \heu in \sectionname~\ref{subsec:overview}, and we detail the algorithms composing our heuristic in \sectionname~\ref{subsec:phases}.

\subsection{Overview\label{subsec:overview}}
At every execution, \heu first considers service requests in decreasing order based on the corresponding service revenue. It then activates the necessary VMs for serving the first service request, trying to map the VNF sequence $w$ onto a path $p$ connecting the VMs deemed to host the required VNFs. 
While doing this, more than one instance can be created for a VNF if necessary to meet the service target delay. To this end, we associate with each VNF a \textit{delay budget}, which is proportional to the VNF computational complexity $\omega(q)$. 
Such budget, however, is {\em flexible}, since the delay contribution of a VNF exceeding its delay budget may be compensated for by a subsequent VNF on $w$, which is deployed in a VM able to process traffic faster than what indicated by the VNF budget. 
Additionally, \heu exploits a {\em backtrack} approach: in case of lack of sufficient resources at a certain point of current path $p$, the algorithm can go back to the last successfully deployed VNF and looks for an alternative deployment (hence path), leaving more spare budget for subsequent VNFs. None the less, it may prove impossible to find enough resources to accommodate the traffic and delay constraint of a given VNF instance; in this case, the service request is rejected.

The decisions that \heu makes are summarized below.

\textit{Placement.} \heu aims to minimize the placement cost. This implies that the number of deployed VNF instances should be low, and the selected VMs should have a low cost. The algorithm thus starts from one instance and chooses the lowest-cost VM among the available ones. If this placement is not feasible, it tries the highest capacity VM to avoid the use of an extra instance. If the latter strategy is also infeasible, it increases the number of instances and repeats the process until a successful deployment is possible, or the limit on the maximum number of instances is reached (\Alg{service-request} and \Alg{placement}). 

\textit{Routing.} Recall that each VNF may have several instances and that such instances may be deployed on VMs connected through multiple logical links. \heu adopts a water-filling approach to route the traffic between each pair of VNFs through different logical links between a pair of VMs. To limit the processing time at each VM, the traffic entering each VM is properly set based on the VM available capacity (\Alg{placement}). 

\textit{CPU assignment.} 
\heu aims to keep the service rate of the used VMs as low as possible, in order to reduce the consumption of computing resources, hence the cost. This means setting the lowest service rate compatible with the per-VNF delay budget, except when we have to compensate for a VNF exceeding its delay budget; in the latter case, the algorithm opts for the maximum service rate on the VM (\Alg{cpu-assignment}). 

\begin{algorithm}[!h]
	\setcounter{AlgoLine}{0}
	\DontPrintSemicolon
	\caption{Main body of \heu algorithm} \label{alg:heuristic-body}
	\KwInput{
		$t,H$, $\mathcal{K}_{t,H} \gets \{k \in \mathcal{K} : \left[t,t+H\right) \cap \big[ t_{\textit{arv}}(k),t_{\textit{dpr}}(k) \big) \neq \emptyset \}$ \label{line:alg1:active-services}
	}
	\KwOutput{
		result sets $\mathcal{R}_p := \{ \mu(k,m,q) \}$, $\mathcal{R}_r := \{ r(k,l,q_1,q_2) \}$, VM states
	}
	
	$\mathcal{R}_p \gets \emptyset ,\mathcal{R}_r \gets \emptyset$
	
	$R(k) \gets X_{\textit{rev}}(s) \cdot ( \min{ \{ t+H, t_{\textit{dpr}}(k) \} } - \max{ \{ t,t_{\textit{arv}}(k) \} } ) \cdot \lambda_{\textit{new}}(s), \forall s \in \mathcal{S}, k \in \mathcal{K}_s \cap \mathcal{K}_{t,H}$ \label{line:alg1:service-revenue}
	
	\textbf{sort} $k \in \mathcal{K}_{t,H} $ \textbf{by} $R(k)$ in desc. order \label{line:alg1:sort}
	
	
	\ForAll{$k \in \mathcal{K}_{t,H}$ \label{line:alg1:for}} {
		\textbf{call} \textsf{BSRD}$(k)$ and update $\mathcal{R}_p$ and $\mathcal{R}_r$ \label{line:alg1:end-for} \;
	}
	\textsf{VM-Activation}$(\mathcal{R}_p)$ \label{line:alg1:vm-activation} \;
\end{algorithm}

\subsection{Algorithms\label{subsec:phases}}

{\bf \Alg{heuristic-body}.} It is the main body of the \heu heuristic, taking as input time horizon $H$, the current time step $t$, and the set $\mathcal{K}_{t, H}$ of service requests which should be served in the time horizon $\left[t,t+H\right)$. \Line{alg1:service-revenue} calculates service revenue $R(k)$ for each request $k$, based on the expected traffic to be served in the time horizon and the expected revenue, \ie, $X_{\textit{rev}}(s)$ for service $s$. The algorithm sorts the service requests in \Line{alg1:sort} in descending order, according to $R(k)$. 
It then calls $\textsf{BSRD}$ for each request, in order to determine whether and how to serve it within the time horizon. If the request can be served, the resulting VNF placement/CPU assignment and routing decisions are stored in $\mathcal{R}_p$ and in $\mathcal{R}_r$, respectively. For each served request, $\mathcal{R}_p$ will then contain a tuple per each VNF instance that specifies the allocated VM and its assigned service rate, while $\mathcal{R}_r$ will contain a tuple for each pair of VNF instances, determining the amount of traffic on their connecting logical link(s). Finally, the VMs required for running the service request are activated if not already active; we recall that it takes one time step to activate them (turning-on state), and they will remain up till the service departure time. 

\begin{algorithm}[!h]
	\setcounter{AlgoLine}{0}
	\DontPrintSemicolon
	\caption{Backtracking-based service request deployment (\textsf{BSRD})}\label{alg:service-request}
	\KwInput{
		service request $k$ of service $s$
	}
	\KwOutput{$\mathcal{R}_p, \mathcal{R}_r$}
	$i \gets 1$; $\textit{status} \gets \tvalue{normal}$; $\textit{can-backtrack} \gets \tvalue{false}$; $\mathcal{C} \gets \emptyset$, $\mathcal{R}_p \gets \emptyset$; $\mathcal{R}_r \gets \emptyset$ \label{line:alg2:init-i} \;
	
	$\Delta(s,q) \gets \omega(q)/\sum_{j=1}^{|\mathcal{Q}_s|}{\omega(Q_s(j))}, \forall q \in$ VNF chain of $s$ \label{line:alg1:delay-budget} \;
	
	\While{$i \le $ number of VNFs \label{line:alg2:while}}	{ 
		
		\If{\textit{status} is \tvalue{normal} \label{line:alg2:if-normal}} {
			
			\For{$n \gets 1$ to $N(s,Q_s(i))$ \label{line:alg2:if-normal1}}{ 
				
				\For{\textit{strategy} $\in \{ \tvalue{cheapest}, \tvalue{largest} \}$ \label{line:alg2:if-normal2}}{ 
					
					\textbf{call} \textsf{VPTR}($k,i,n,\textit{strategy}$) and \textsf{CA}($k,i$)\label{line:alg2:call-normal} \;
					
					\lIf{deployment is successful} {
						\textbf{break}\label{line:alg2:endif-normal}
					}
				}
			}
		}\ElseIf{\textit{status} is \tvalue{critical} \label{line:alg2:if-critical}} { 
			$\textit{can-backtrack} \gets \tvalue{false}$ \label{line:alg2:if-critical1} \;
			\textbf{call} \textsf{VPTR}($k,i,N(s,Q_s(i)),\tvalue{largest}$) and \textsf{CA}($k,i$) \label{line:alg2:endif-critical} \;
		}	
		\If{$i$-th VNF is successfully deployed \label{line:alg2:if-place-success}} { 
			\lIf{\textit{status} is \tvalue{normal}} {
				\textit{can-backtrack} $\gets$ \tvalue{true} \label{line:alg2:if-place-success1}
			}
			
			Update $\mathcal{R}_p, \mathcal{R}_r, \textit{status} \gets \tvalue{normal}, i\gets i+1$ 	\label{line:alg2:if-place-success2} \;
		} \Else (\Comment*[f]{$i$-th VNF is not deployed}) { \label{line:alg2:else-place-success}
			\textit{status} $\gets$ \tvalue{critical}\;
			\If{\textit{can-backtrack} \label{line:alg2:else-place-success1}}{ 
				Discard $\mathcal{R}_p, \mathcal{R}_r$ for $(i-1)$-th VNF, $i\gets i-1$ 	\label{line:alg2:else-place-success2} \;
			} \ElseIf{fail is due to delay budget \label{line:alg2:else-place-success3}}{
				Update $\mathcal{R}_p, \mathcal{R}_r$, $i\gets i+1$ \label{line:alg2:else-place-success4} \;
			}\Else (\Comment*[f]{fail is due to traffic}) { \label{line:alg2:else-place-success5}
				\textbf{terminate} and discard $\mathcal{R}_p$ and $\mathcal{R}_r$ \label{line:alg2:else-place-success6}
				\label{line:alg2:endif-place-success} \;
			}
		}
		\If{result sets are not feasible \label{line:alg2:if-feasible}} { 
			\textbf{terminate} and discard $\mathcal{R}_p$ and $\mathcal{R}_r$ \;
		}	
	}
\end{algorithm}

{\bf \Alg{service-request}.} Given service request $k$ for service $s$ as an input, the goal of \Alg{service-request} is to check whether all VNFs of $s$ can be deployed with the available resources. If it is possible, the request is served and the result sets $\mathcal{R}_p$ and $\mathcal{R}_r$ are returned. The global boolean variables \textit{status} and \textit{can-backtrack} represent the deployment status and the possibility of backtracking, respectively. \textit{status} is \tvalue{critical} if the last VNF deployment has failed, and \tvalue{normal} otherwise. The global cache $\mathcal{C}$ is a set of results that facilitates the backtracking operation (see \Alg{placement}). The algorithm starts in \tvalue{normal} mode; clearly, backtracking is not allowed for the first VNF in the VNFFG and cache $\mathcal{C}$ is empty (\Line{alg2:init-i}). 
The algorithm starts by assigning a delay budget to each VNF of the service, which is proportional to the VNF computational complexity (\Line{alg1:delay-budget}), where $Q_s(j)$ denotes the $j$-th VNF in the VNFFG. Then, it goes across the sequence of VNFs starting from the ingress VNF and deploys them one by one. 

For each VNF, \mLine{alg2:if-normal}{alg2:endif-critical} decide on the number of required instances and the VM selection \textit{strategy}, based on the deployment \textit{status}. The strategy can be \tvalue{cheapest} or \tvalue{largest}: the algorithm selects VMs with the lowest cost when the strategy is \tvalue{cheapest}, and with the highest capacity when the strategy is \tvalue{largest}. 
The first part (\mLine{alg2:if-normal1}{alg2:endif-normal}) deploys the VNF in the \tvalue{normal} mode. Since the algorithm aims to keep the number of required VNF instances as low as possible, it starts with one instance and the \tvalue{cheapest} strategy and calls $\textsf{VPTR}$ to determine placement and routing, and $\textsf{CA}$ to determine the CPU assignment. The deployment is successful if neither of these algorithms fails. If the \tvalue{cheapest} strategy does not yield a successful deployment for the VNF, the algorithm keeps the number of instances fixed and tries the \tvalue{largest} strategy. If both strategies fail, the number of instances is increased by one and the process is repeated. The algorithm ends whenever a successful deployment is found (\Line{alg2:endif-normal}), or the maximum number of instances is reached.

\mLine{alg2:if-place-success}{alg2:endif-place-success} decide how to proceed in the VNF sequence according to the result of deployment, \textit{status} and \textit{can-bakctrack}. If the deployment is successful (\Line{alg2:if-place-success}), the algorithm updates the result set, sets \textit{status} to \tvalue{normal} and proceeds to the next VNF in the VNFFG (\Line{alg2:if-place-success2}). \textit{can-backtrack} is also updated in \Line{alg2:if-place-success1}, which means that backtracking is allowed for the next VNFs only when we have a successful deployment in the \tvalue{normal} mode for the current VNF: this prevents the algorithm to backtrack again to a VNF, which has already been deployed in \tvalue{critical} mode. 
Otherwise (\Line{alg2:else-place-success}), \textit{status} is set to \tvalue{critical} and the algorithm proceeds as follows. As the first attempt, it tries to refine the placement in the previous step. Thus, if backtracking is allowed, it reverts the result sets related to the previous VNF in the VNFFG and goes back to deploy it again (\Line{alg2:else-place-success2}). When the deployment fails but backtracking is not possible, due to a violation of the delay budget, the algorithm preserves the current deployment in the result set and proceeds to the next VNF, hoping to compensate for the exceeded delay budget (\Line{alg2:else-place-success4}). If neither option is viable, the algorithm decides not to serve the current service request and reverts all result sets related to its deployment (\Line{alg2:endif-place-success}).

\mLine{alg2:if-critical1}{alg2:endif-critical} deploy the VNF when status is \tvalue{critical}, \ie, when the previous VNF deployment has failed. This VNF is either the next VNF in the VNFFG when the algorithm is in the backtracking phase, or the previous VNF when the algorithm is going to compensate for the exceeded delay budget by the current deployment. In either case, the algorithm chooses the fastest option to deploy the VNF, regardless of the cost, using the maximum number of instances and \tvalue{largest} strategy. 
Finally, the algorithm checks the feasibility of the decisions made with regard to the datacenter capacity and service target delay after each VNF deployment in \Line{alg2:if-feasible}. For the former, it is enough to check that the total computational capability assigned to VMs within each datacenter does not exceed its maximum capacity, \ie, for each datacenter $d$,
\begin{equation}
\sum_{\substack{\mu(k,m,q) \in \mathcal{R}_p : m \in \mathcal{M}_d}}{\mu(k,m,q) \cdot \omega(q)} \le C_{\textit{dc}}(d).
\end{equation} 
Traffic packets belonging to a service may go through different end-to-end paths in the physical network and experience different end-to-end delays. We define $\bar{\delta}(k,m,q)$ as the maximum end-to-end delay that traffic packets belonging to service request $k$ experience from the ingress VNF until they depart VM $m$ which hosts an instance of VNF $q$. Thus, after deploying VNF $q$ of service request $k \in \mathcal {K}_s$, it is enough to check that this delay for any VM $m$, hosting an instance of $q$, does not exceed the service target delay:
\begin{equation}
\bar{\delta}(k,m,q) \le D_{\textit{QoS}}(s).
\end{equation}

\begin{algorithm}[!h]
	\DontPrintSemicolon
	\setcounter{AlgoLine}{0}
	\caption{VNF placement and traffic routing (\textsf{VPTR})}\label{alg:placement}
	\KwInput{$k \in \mathcal{K}_s,i,n,\textit{strategy}$}
	$(q_1,q_2) \gets (Q_s(i-1),Q_s(i));$ $\mathcal{R}_r \gets \emptyset;$ $\Lambda' \gets \Lambda(s,q_1,q_2);$ \label{line:alg3:init-vnf} \;
	
	$B_{\textit{log}}'(l) \gets $ remaining capacity of $l, \forall l \in \mathcal{L}$ \label{line:alg3:init-cl} \;
	
	$\mathcal{L}' \gets \{ l \in \mathcal{L} \cup \mathcal{L}_{\circ} : q_1 \text{ is on } \textit{src(l)} \wedge \textit{dst(l)} \text{ is free} \wedge	B_{\textit{log}}'(l)>0\}$\label{line:alg3:init-L}\;
	
	\If(\Comment*[f]{cache is not empty}){$\mathcal{C} \neq \emptyset$ \label{line:alg3:cache1}}{ 
		
		Fill $l \in \mathcal{L}' : \textit{dst}(l)=m$, consiedring limit $\frac{D(k,m,q)}{\alpha(s,q)}$, $\forall D(k,m,q) \in \mathcal{C} : q=q_2$
		
		Update $\mathcal{R}_r, \Lambda',n,\mathcal{L}';$ $\mathcal{C} \gets \emptyset$ \label{line:alg3:cache3} \;
		
	}
	\If{\textit{strategy} is \tvalue{cheapest}} {
		\textbf{sort} $l \in \mathcal{L}'$ \textbf{by} $\omega(q_2) \cdot X_{\textit{cpu}}(\textit{dst(l)}) + \sum_{e \in l}{X_{\textit{link}}(e)}$ in asc. order \label{line:alg3:sort-cheapest}
	} \ElseIf{\textit{strategy} is \tvalue{largest}} {
		\textbf{sort} $l \in \mathcal{L}'$ \textbf{by} $\min{ \{ B_{\textit{log}}'(l), \frac{C_{\textit{vm}}(\textit{(dst(l))}}{\omega(q_2)} \} }$ 
		in desc. order \label{line:alg3:sort-largest}
	}
	$\mathcal{L}_{\textit{top}} \gets $ Pick top $l \in \mathcal{L}'$ as much as possible such that $|\{ \textit{dst(l)} : l\in\mathcal{L}_{\textit{top}} \}|=n$ \label{line:alg3:init-L-top}
	
	$\mathcal{M}_{\textit{top}} \gets \{ \textit{dst(l)} : l\in \mathcal{L}_{\textit{top}} \}$ \label{line:alg3:init-M-top}\;
	
	$\hat{I}(k,m,q_2) \gets \frac{C_{\textit{vm}}(m)}{\sum_{m'\in\mathcal{M}_{\textit{top}}}{C_{\textit{vm}}(m')}} \cdot \Lambda', \forall m \in \mathcal{M}_{\textit{top}}$ 
	\label{line:alg3:proportional}
	
	$C_{\textit{vm}}'(m) \gets C_{\textit{vm}}(m),\forall m \in \mathcal{M}_{\textit{top}}$
	\label{line:alg3:init-cm}
	
	\ForAll{$l\in\mathcal{L}_{\textit{top}}$ \label{line:alg3:water-filling1}}{ 
		$c(l) \gets \min \{ B_{\textit{log}}'(l), \frac{C_{\textit{vm}}'(\textit{dst(l)})}{\omega(q_2)} \}$ \label{line:alg3:water-filling2}
		
		$r(k,l,q_1,q_2) \gets $ Fill $l$ by remaining outgoing traffic of $q_1$ on \textit{src(l)} considering $c(l)$ and limit $\hat{I}(k,\textit{dst(l)},q_2)$ \label{line:alg3:water-filling3}
		
		Update $\Lambda'$, $B_{\textit{log}}'(l)$, $C_{\textit{vm}}'(\textit{dst(l)});$ $\mathcal{R}_r \gets \mathcal{R}_r \cup \{ r(k,l,q_1,q_2) \}$ \label{line:alg3:water-filling4} \label{line:alg3:water-filling5}
	}
	\If{$\Lambda'>0$ \label{line:alg3:if-lambda1}}{
		Preserve $D(k,m,q_1)$ in cache $\mathcal{C}$ such that $q_1$ is on $m$ \label{line:alg3:if-lambda2} \;
		
		\Return{\tvalue{fail}, $\emptyset$ \label{line:alg3:if-lambda3}}
	}
	\Return{\tvalue{success}, $\mathcal{R}_r$ \label{line:alg3:success}}
\end{algorithm}

{\bf \Alg{placement}.} It determines the placement and traffic routing for the $i$-th VNF of request $k$ of service $s$, using $n$ instances and the given \textit{strategy}. \Line{alg3:init-vnf} initializes $(q_1,q_2)$ to the $i$-th VNFs pair in the VNFFG of service $s$, the routing result set $\mathcal{R}_r$ to $\emptyset$, and the remaining unserved traffic between $q_1$ and $q_2$, \ie, $\Lambda'$, to $\Lambda(k,q_1,q_2)$. The first pair of VNFs is $(\circ, q_1)$ with the assumption that the \textit{dummy} VNF $\circ$ is placed on the \textit{dummy} VM. 
In \mLine{alg3:init-cl}{alg3:init-L}, first the remaining capacity of each logical link $l$ is calculated and stored in $B_{\textit{log}}'(l)$ and then the ones that have a remaining capacity greater than zero, host VNF $q_1$ on their source VM, and host no VNF on their destination, are picked and stored in the set $\mathcal{L}_{\textit{top}}$. The links in $\mathcal{L}_{\textit{top}}$ and their destination VMs are the only potential candidates for this algorithm to place instances of the $i$-th VNF and accommodate its incoming traffic $\Lambda'$. In other words, in the rest of the algorithm, we consider the joint logical link and its destination VM as one entity and pick the best ones according to the \textit{strategy} and $n$. If the selected entities cannot fit the incoming traffic, the placement fails; none the less, we still preserve the amount of satisfied traffic in the cache and exploit this information in the backtracking phase.

The implementation speed of the backtrack operation is greatly improved by {\em caching}. Specifically, when \Alg{placement} is called in the backtracking phase to refine the placement of $i$-th VNF, the cache contains results which determine the routing of a portion of the outgoing traffic of the $(i+1)$-th VNF to the $(i+2)$-th VNF, which was satisfied by the previous deployment of the $(i+1)$-th VNF in the VNFFG. \mLine{alg3:cache1}{alg3:cache3} exploit the cached results and accommodate the unserved portion of incoming traffic by using different instances, which helps the next deployment of the $(i+1)$-th VNF to fully serve its traffic.
For instance, assuming $\alpha(s, \mathcal{Q}_s(i+1))=1$ and that the placement of the $(i+1)$-th VNF has failed by $\Lambda'$ unserved traffic, the backtracking step will have to accommodate only $\Lambda'$ traffic on extra VMs, \ie, the routing and placement results for the served traffic portion, $D(k, m, q_2) \in \mathcal{C}$, will not change. 

The pairs of logical links and the connected VMs will be selected for placement and routing based on the given \textit{strategy}. If the \textit{strategy} is \tvalue{cheapest}, they will be sorted according to the cost of the logical link plus the VM CPU cost in ascending order (\Line{alg3:sort-cheapest}). If the strategy is \tvalue{largest}, we sort them in descending order by the minimum of the remaining capacity of the logical link and the VM (\Line{alg3:sort-largest}). 
\Line{alg3:init-L-top} picks the biggest set of top logical links such that the number of unique destination VMs is equal to the number of instances, \ie, $n$, and stores them in $\mathcal{L}_{\text{top}}$. Note that there may be multiple logical links with the same destination VM in this set, and therefore we should pick the largest set to increase the chance of fitting the traffic. If the number of unique destination VMs is less than $n$, $\mathcal{L}_{\text{top}}$ will be empty and the placement fails. Otherwise, we store destination VMs corresponding to logical links $l \in \mathcal{L}_{\text{top}}$ in set $\mathcal{M}_{\text{top}}$ (\Line{alg3:init-M-top}).

To avoid an exceedingly high processing time, \Line{alg3:proportional} introduces a limit for the amount of traffic entering a given VM $m \in \mathcal{M}_{\text{top}}$, proportional to the VM maximum computational capacity. Notice that all logical links ending at the same destination VM have the same limit. The remaining computational capacity of each selected VM, $C'_{vm}(m)$, is initialized to its maximum $C_{vm}(m)$ (\Line{alg3:init-cm}). The algorithm adopts a \textit{water-filling} approach to fill the logical links in \mLine{alg3:water-filling1}{alg3:water-filling4}. First, for each logical link $l$ and its connected VM $dst(l)$, the remaining capacity, \ie, the minimum of the remaining capacities of $l$ and $\textit{dst(l)}$, is stored in $c(l)$ (\Line{alg3:water-filling2}). Then, logical link $l$ is filled by the remaining unserved outgoing traffic of VNF $q_1$ on VM $\text{src}(l)$, so that neither $c(l)$ limit on the capacity of logical link $l$ nor the $\hat{I}(k,m,q_2)$ limit on the incoming traffic of VM $\text{dst}(l)$ are violated. \Line{alg3:water-filling4} updates the remaining unserved traffic from $q_1$ to $q_2$ ($\Lambda'$), the remaining capacity of logical link $l$ ($B_{\text{log}}'(l)$), the remaining capacity of destination VM ($C_{\textit{vm}}'(\textit{dst}(l))$), and routing result set ($\mathcal{R}_r$). 
Finally, if there is still some unserved traffic from VNF $q_1$ to $q_2$ (\ie, not all the traffic can be served), the algorithm returns \tvalue{fail} (\mLine{alg3:if-lambda1}{alg3:if-lambda3}). \Line{alg3:if-lambda2} preserves the satisfied outgoing traffic of VM $m$ hosting an instance of VNF $q_1$, \ie, $D(k,m,q_1)$, in the cache, so as to use it later on in case of backtracking. Otherwise, the algorithm returns \tvalue{success} with the placement result set $\mathcal{R}_p$.

\begin{algorithm}[!h]
	\setcounter{AlgoLine}{0}
	\DontPrintSemicolon
	\caption{CPU assignment (\textsf{CA})}\label{alg:cpu-assignment}
	\KwInput{$k \in \mathcal{K}_s,i, \mathcal{R}_r$
}
	
	$(q_1,q_2) \gets (Q_s(i-1),Q_s(i));$ $\mathcal{R}_p \gets \emptyset;$ \label{line:alg4:init-vnf} \;
	
	$\mathcal{L}_{\textit{dep}} \gets \{ l \in\mathcal{L} : \exists r(k',q'_1,q'_2,l) \in \mathcal{R}_r : k'=k \wedge q'_1=q_1 \wedge q'_2=q_2 \wedge r(k',q'_1,q'_2,l)>0\}$ \label{line:alg4:init-lm} \;
	
	$\mathcal{M}_{\textit{dep}} \gets \{ m \in\mathcal{M} : \exists l \in \mathcal{L}_{\textit{dep}} : \textit{dst}(l)=m \}$ \label{line:alg4:init-m-dep} \;
	
	\For{$m \in \mathcal{M}_{\textit{dep}}$ \label{line:alg4:normal1}}{ 
		$I(k,m,q_2) \gets \sum_{\substack{r(k,l,q_1,q_2) \in \mathcal{R}_r : \textit{dst}(l)=m}}{r(k,l,q_1,q_2)}$ \label{line:alg4:init-Im} \;
		
		$\check{\delta}(k,m,q_2) \gets \max\limits_{\substack{l \in \mathcal{L}_{\textit{dep}} :\textit{dst}(l)=m } }{ \left( \bar{\delta}(k, \textit{src(l)}, q_1) + D_{\textit{log}}(l) \right) }$ \label{line:alg4:init-deltam} \;
		
		\If{\textit{status} is \tvalue{critical}}{
			$\mu(k,m,q_2) \gets \frac{C_{\textit{vm}}(m)}{\omega(q_2)}$ \label{line:alg4:critical} \;
		}\Else(\Comment*[f]{\textit{status} is \tvalue{normal}}) {
			
			$\mu(k,m,q_2) \gets I(k,m,q_2) + \frac{1}{ \sum_{j=1}^{i}{\Delta(s,Q_s(j))} - \check{\delta}(k,m,q_2)}$ \label{line:alg4:normal2} \;
			
			\If{$\mu(k,m,q_2) \notin (I(k,m,q_2), \frac{C_{\textit{vm}}(m)}{\omega(q_2)}]$ \label{line:alg4:normal3}}{ 
				$\mu(k,m,q_2) \gets \frac{C_{\textit{vm}}(m)}{\omega(q_2)}$ \label{line:alg4:normal4}\;
			}
		}
		
		$\mathcal{R}_p \gets \mathcal{R}_p \cup \{ \mu(k,m,q_2) \}$ \label{line:alg4:update-set} \;
		$\bar{\delta}(k, m,q_2) \gets \check{\delta}(k,m,q_2) + \frac{1}{\mu(k,m,q_2) - I(k,m,q_2) }$ \label{line:alg4:update-delta} \;
		
	}

	\If{$\max_{m \in \mathcal{M}_{\textit{dep}}}{\bar{\delta}(k, m,q_2)} > \sum_{j=1}^{i}\Delta(s,Q_s(j))$ \label{line:alg4:delay-contribution}}{
		\Return{\tvalue{fail}, $\mathcal{R}_p$}
		
	}
	\Return{\tvalue{success}, $\mathcal{R}_p$}
\end{algorithm}

{\bf \Alg{cpu-assignment}}. It is called in \Line{alg2:call-normal} and \Line{alg2:endif-critical} of \Alg{service-request} when the deployment of VNF $q$ in \Alg{placement} is successful. Given the result set $\mathcal{R}_r$, this algorithm is responsible for assigning the service rates to VMs for running the deployed instances of VNF $q$. After initialization, in \Line{alg4:init-lm}, $\mathcal{L}_{\textit{dep}}$ defines the set of the logical links used for routing a part of traffic from any instance of VNF $q_1$ to any instance of VNF $q_2$. We store the VMs on which VNF $q_2$ is already deployed in the set $\mathcal{M}_{\textit{dep}}$ (\Line{alg4:init-m-dep}). Then, for each $m \in \mathcal{M}_{\textit{dep}}$, we calculate the incoming traffic through the sum of traffic from all logical links ending in VM $m$, and store it in $I(k,m,q_2)$ in \Line{alg4:init-Im}.

$\check{\delta}(k,m,q_2)$ represents the maximum end-to-end delay that traffic packets experience from the ingress VM to VM $m$, which hosts an instance of VNF $q_2$, but before being processed by $m$. For each logical link $l \in \mathcal {L}_{\textit{dep}}$ where $\textit{dst(l)}=m$, this delay is equal to the sum of the maximum end-to-end delay of traffic packets after being processed by VNF $q_1$ on VM $\textit{src}(l)$, \ie, $\bar{\delta}(q_1,\textit{src(l)})$, and the delay of logical link $l$, \ie, $D_{\textit{log}}(l)$. Taking the maximum over all such logical links, we have $\check{\delta}(k,m,q_2)$ in \Line{alg4:init-deltam}.

Similar to the VNF deployment in \Alg{placement}, the algorithm assigns service rates to VMs based on the deployment \textit{status}. In the \tvalue{critical} mode, the algorithm aims to reduce the delay contribution, which depends on logical links delay and processing time on VMs. The logical links are already selected by the \textsf{VPTR} algorithm, thus here we assign the maximum possible service rate for the VM to reduce the processing time (\Line{alg4:critical}). 
Instead, when the algorithm is in \tvalue{normal} mode, it chooses the minimum possible service rates for VM $m$ (\Line{alg4:normal2}), such that the VNFs delay budget do not violate, \ie \begin{equation}\label{eq:miu-delay}
{ \sum_{j=1}^{i}{\Delta(s,Q_s(j))} - \check{\delta}(k,m,q_2)} = \frac{1}{ \mu(k,m,q_2) - I(k,m,q_2) }.
\end{equation}
In the above equation, the right- and left-hand sides represent the processing time of VM $m$ and the remaining delay budget of VNFs, respectively. To compute the latter, first it is calculated the total delay budget of the VNFs up to the $i$-th one (\ie, the current one). Then, it is subtracted by the maximum end-to-end delay of traffic packets, before being processed by VNF $q_2$ on VM $m$, \ie, $\check{\delta}(k,m,q_2)$.

The computed service rate for VM $m$ may be invalid because (i) 
no delay budget is left to process the current VNF on VM $m$, \ie, the left-hand side of equality in \Eq{miu-delay} becomes non-positive, or (ii) the assigned service rate exceeds the maximum capability of the VM. 
In both cases, the \textsf{CA} algorithm fails, however the VM is assigned to its maximum computational capability to process the VNF (\Line{alg4:normal4}). Recall that, although the CPU assignment failed for the current VNF, the algorithm keeps the results to be used in \Alg{service-request} (\Line{alg2:else-place-success3}) when backtracking is not allowed. In this case, the algorithm continues with the next VNF and tries to compensate for the exceeded delay budget. \Line{alg4:update-set} stores the results, and \Line{alg4:update-delta} updates $\bar{\delta}(k,m,q_2)$ for this VM that shows the maximum end-to-end delay after the packets are processed by VM $m$. 
Finally, when all service rates have been assigned, the algorithm returns \tvalue{fail} if the remaining delay budget is violated for at least one VM (\Line{alg4:delay-contribution}), and \tvalue{success} otherwise.

\subsection{Computational Complexity}
The \heu heuristic takes the set of physical links $\mathcal {E}$, service requests $\mathcal {K}$, and their VNFFG $\mathcal{Q}_s$, VMs $\mathcal {M}$, and logical links $\mathcal{L}$ as inputs. Note that $\mathcal{L}$ is considered as an input since it is computed once for all executions of \heu algorithm. Below, we prove that this algorithm has a worst-case polynomial complexity in terms of input parameters.
\begin{theorem}
	The \heu algorithm has a worst-case polynomial computation complexity.
\end{theorem}

\begin{IEEEproof}
	First, we determine the complexity of the \textsf{VPTR} and \textsf{CA} algorithms. \textsf{VPTR} constructs and sorts the set $\mathcal{L}'$ in $O(|\mathcal{L}| \log{|\mathcal{L}|})$ and adopts water filling to fill the logical links in $O(\mathcal L)$, thus the total time complexity of this algorithm is $O(|\mathcal{L}| \log{|\mathcal{L}|})$. \textsf{CA} also has $O(\mathcal L)$ complexity, hence the total computational complexity of \textsf{VPTR} and \textsf{CA} remains equal to that of \textsf{VPTR}. 
	\Alg{heuristic-body} sorts the service requests in $O(|\mathcal{K}| \log{|\mathcal{K}|})$ and calls \textsf{BSRD} for each service request. In the worst-case, \textsf{BSRD} tries every possible number of instances and strategies for all VNFs in the VNFFG of the given service request. Let $N$ and $Q$ be upper bounds on the maximum number of instances, \ie $N(s, q), \forall s \in \mathcal{S}, q \in \mathcal{Q}$, and the number of VNFs in a VNFFG, \ie $|\mathcal Q_s|, \forall s \in \mathcal{S}$, respectively. Thus, the total time complexity of \textsf{BSRD} is $O(N Q |\mathcal{L}| \log{|\mathcal{L}|})$ and total time complexity of~\Alg{heuristic-body} is $O\big(|\mathcal{ K}| N Q |\mathcal{L}| \log{|\mathcal{L}| } + |\mathcal{K}| \log{|\mathcal{K}|} \big)$.	
	Therefore, the worst-case total time complexity is polynomial in terms of input parameters. In other words, the complexity of the heuristic depends primarily on the number of service requests, the number of VNFs in each VNFFG, the number of deployment attempts for each VNF, and the number of logical links.
\end{IEEEproof}

\section{Numerical Results\label{sec:numerical-results}}

We now present the results of the numerical experiments we conducted, and show that our proposed scheme consistently performs better than state-of-the-art approaches and close to the optimum. 
We compare our heuristic algorithm against the following benchmarks:
\begin{itemize}
	\item \textbf{Global optimum}. 
	The solution of the optimization problem defined in \sectionname~\ref{sec:problem-formulation} obtained by brute-force search, assuming exact knowledge of arrival and departure times of all service requests.
	\item \textbf{Best-fit}. It is an online algorithm which decides about each service request upon its arrival, without any information about the future service requests. \bestfit deploys VNFs of a service request one by one, using a single instance of each VNF and the \tvalue{cheapest} strategy. If the request can be served, the selected resources will be dedicated to the service request until its departure.
\end{itemize}
In our performance evaluation, we use the following performance metrics:
\begin{itemize}
	\item \textbf{Service revenue}, defined as the sum of revenues achieved by serving service requests. For a single request of service $s$, this metric equals the amount of served traffic multiplied by $X_{\textit{rev}}(s)$.
	\item \textbf{Cost/traffic}, which reflects the average cost incurred to serve a unit of traffic. 
\end{itemize} 

In the following, we first consider a small-scale network scenario, for which the optimum solution can be obtained in a reasonable time. This scenario will give interesting and easy-to-interpret insights regarding how each service type impacts the revenue and \ratio. Then, we run \heu and \bestfit in a large-scale real network scenario, where achieving the optimum solution is impractical. 
\Tab{services} summarizes the services we consider for our performance evaluation, inspired to real-world 5G applications. The revenue gained from serving one unit of traffic of service $s$, \ie, $X_{\textit{rev}}(s)$, is set inversely proportional to the service target delays. We assume that the service requests arrive according to a Poisson process, and the duration of requests follows an exponential distribution. 

\newcolumntype{L}{>{\raggedright\arraybackslash}X}
\begin{table}[t]
	\caption{List of services}
	\label{table:services}
	\centering
	\begin{tabularx}{0.5\textwidth}{@{}ccccL@{}}
		\toprule 
		Service & $D_{\textit{QoS}}$ & $\lambda_{\textit{new}}$ & $X_{\textit{rev}}$ & Application \\ 
		& (ms) & (Mb/s) & (\euro/Gb) & \\ 
		\midrule
		$s_1$ & $10$ & $3$ & $100$ & safety apps. (\eg, vehicle collision detection) \\ 
		$s_2$ & $45$ & $10$ & $22.2$ & real-time apps (\eg, gaming) \\ 
		$s_3$ & $80$ & $15$ & $12.5$ & soft real-time apps \\ 
		$s_4$ & $2500$ & $400$ & $0.4$ & delay-tolerant apps (\eg, video streaming)\\ 
		\bottomrule
	\end{tabularx}
\end{table}

In both scenarios, we study the impact of traffic and delay on the performance metrics by multiplying traffic arrival rates $\lambda_{\textit{new}}$ and physical link delays $D_{\textit{phy}} (e)$ by different factors. We run each experiment $50$ times and report the average value for each point in the figures. 
In general, \heu, taking advantage of backtracking, achieves close to the optimum service revenue better than \bestfit. However, the value of \ratio depends on how tight the target delay is. When the target delay is small, the chance of backtracking increases; therefore, \heu incurs more cost to serve the requests.

\subsection{Small-scale Scenario}
We consider two pairs of VMs of different types, \ie, \textit{small} and \textit{medium} as described in \Tab{vm-types}. 
Pairs of VMs inside are connected using a physical link: physical links between \textit{small} and \textit{medium} types VMs have cost of $0.02$~\euro/Gb and $0.04$~\euro/Gb per hour, respectively, while their latency varies from $1$~ms to $7$~ms with the default value set to $2$~ms, and we disregard the link capacity. The time needed to setup a VM is one minute.

\begin{table}[t]
	\caption{Different VM types in datacenters}
	\label{table:vm-types}
	\centering
	\begin{tabular}{ccccc}
		\toprule
		VM type & $C_{vm}$ & $X_{cpu}$ & $X_{idle}$ \\ 
		& (MIPS) & (\euro/MIPS/hour) & (\euro/hour) \\ 
		\midrule
		Small & $600$ & $2 \times 10^{-5}$ & $0.018$ \\ 
		Medium & $1200$ & $4 \times 10^{-5}$ & $0.036$ \\
		Large & $1800$ & $6 \times 10^{-5}$ & $0.054$ \\
		\bottomrule
	\end{tabular}
\end{table}

We consider two simple services $s_1$ and $s_2$, each having a chain of two VNFs with target delays $10$~ms and $45$~ms, and with input traffic rates $3$~Mb/s and $15$~Mb/s, respectively (as summarized in \Tab{services}). In this scenario, we set $N(s, q)=1$ for all VNFs, an average duration of $3$ minutes for each service, and we assign them randomly to the arrival points of a Poisson process with an average rate of $0.5$ requests per minute, while the total system lifespan is set to $10$ minutes.

\begin{figure}[t]
	\centering
	\subfloat[\label{fig:lambda-small-rev}]{{\includegraphics[width=0.5\linewidth]{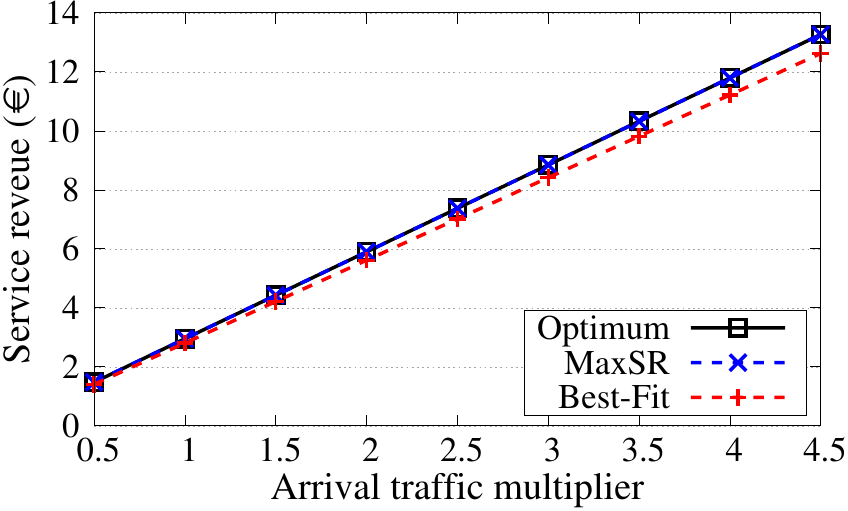} }}
	\subfloat[\label{fig:lambda-small-cost}]{{\includegraphics[width=0.5\linewidth]{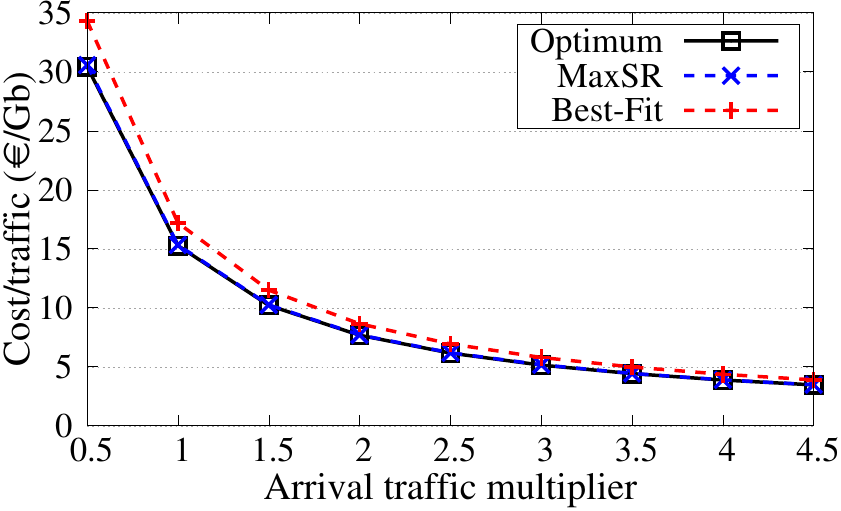} }}
	\caption{Small-scale scenario. Impact of service requests arrival traffic on absolute value of service revenue and \ratio. Physical link delay = $2$~ms.}
	\label{fig:lambda-small}
\end{figure}

{\bf Impact of Physical Link latency and Arrival Traffic.}
\Fig{lambda-small} shows the impact of the traffic arrival intensity on the service revenue and \ratio. \heu matches the optimum, and \bestfit performs close to the optimum in both service revenue and \ratio. As it has no backtracking mechanism, \bestfit does not serve a request whenever any of its VNFs cannot be served within its delay budget, \ie, it has no budget flexibility; therefore, it achieves lower service revenue than the optimum. While the cost of physical links increases proportionally to the traffic, the costs of VMs in \textit{turning-on} mode remains constant, and their cost in \textit{active} mode increases less than proportionally with the traffic; the resulting effect is that \ratio decreases with the traffic -- which conforms to the intuitive notion that serving larger amounts of traffic is more cost-efficient. \bestfit incurs more cost compared to \heu and optimum because it does not support VNF migration, causing a VNF to continue running on a high-cost VM even if a low-cost VM becomes available. The excess VMs CPU cost and transmission cost scale with the traffic, whereas the excess VMs idle cost remains constant; therefore, the difference between \bestfit and optimum becomes smaller as traffic increases.

\begin{figure}[t]
	\centering
	\subfloat[\label{fig:delay-small-rev}]{{\includegraphics[width=0.5\linewidth]{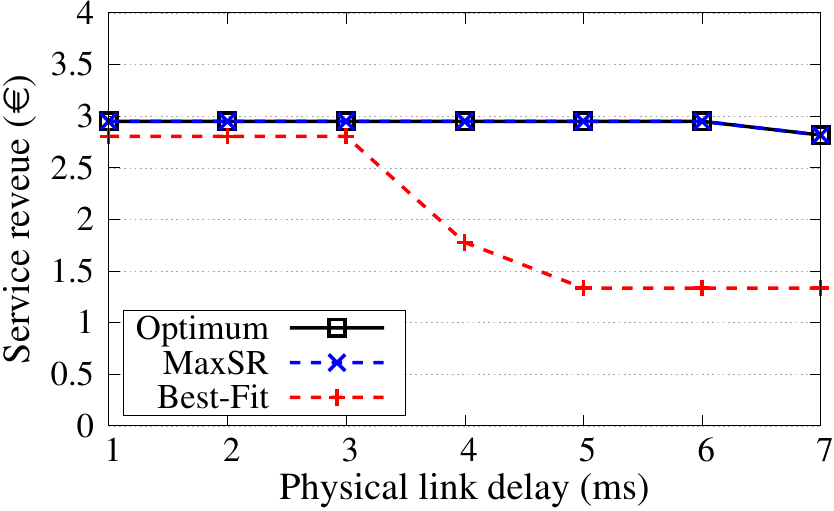} }}
	\subfloat[\label{fig:delay-small-cost}]{{\includegraphics[width=0.5\linewidth]{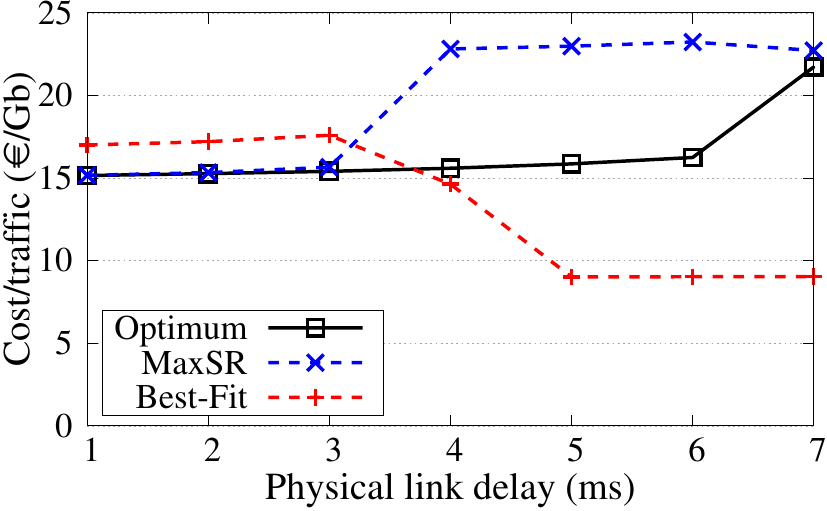} }}
	\caption{Small-scale scenario. Impact of physical link latecy on absolute value of service revenue and \ratio. Arrival traffic multiplier = $1$.}
	\label{fig:delay-small}
\end{figure}
\begin{figure}[t]
	\centering
	\subfloat[Physical link delay = $3$~ms. \label{fig:delay-small-service1}]{{\includegraphics[width=0.5\linewidth]{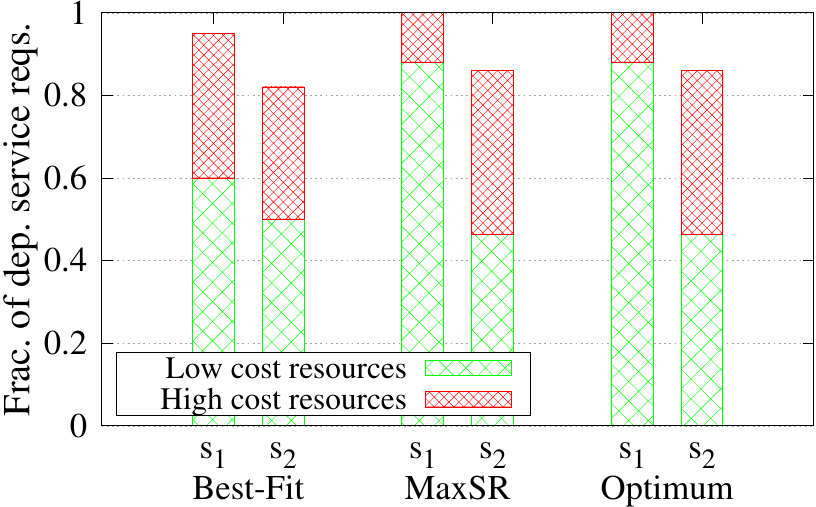} }}
	\subfloat[Phsysical link delay = $7$~ms.\label{fig:delay-small-service2}]{{\includegraphics[width=0.5\linewidth]{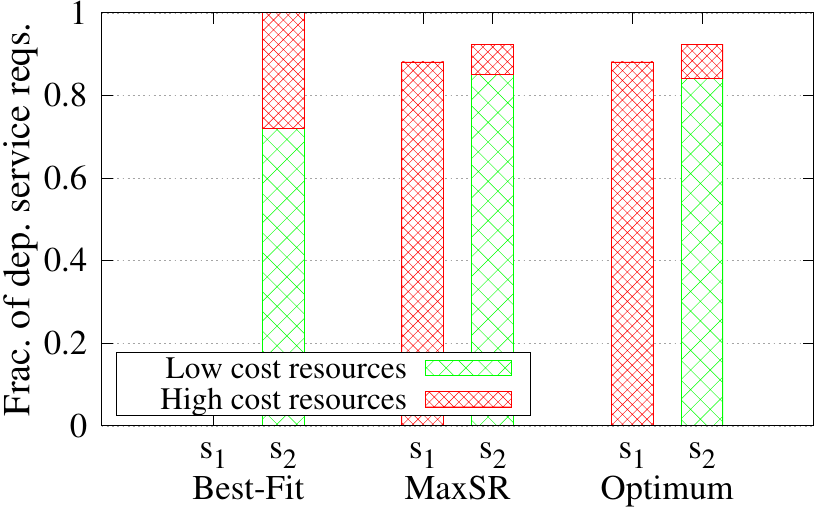} }}
	\caption{Fraction of deployed service requests for each service and algorithm. Arrival traffic multiplier = 1.}
	\label{fig:delay-small-service}
\end{figure}

\Fig{delay-small} shows the impact of physical link latency on the service revenue and cost/traffic ratio. For all latency values, \heu is still able to achieve optimum service revenue. As shown in~\Fig{delay-small-service1}, no strategy (not even optimum) can serve all requests when the physical link delay is $3$~ms especially, for service $s_2$. The reason is that when the number of concurrent requests becomes more than two, both optimum and \heu give the priority to the high-revenue service $s_1$ and requests for $s_2$ will only be processed if resources are available. When the physical link delay increases, service requests need more computational capacity on VMs to meet their target delay, in order to offset longer network delays. Specifically, when the physical link delay is $7$~ms, requests of type $s_1$ can only be served on high-capacity VMs, and therefore, concurrent requests of this type can not be served. This is confirmed by the degradation of the optimum in~\Fig{delay-small-rev}, and in~\Fig{delay-small-service2}, where the fraction of served requests of type $s_1$ becomes less than $1$ when the physical link delay is $7$~ms. 

\bestfit gains substantially lower service revenue compared to others, especially for higher values of physical link delay. As shown in \Fig{delay-small-service2}, this is due to the fact that \bestfit cannot deploy requests of type $s_1$ in those cases. This, in turn, is due to the fact that it does not support backtracking: when the delay budget for the second VNF in the chain of $s_1$ is violated, no corrective action is taken and the whole request fails.

\Fig{delay-small-cost} shows \heu has a higher \ratio when the physical link delay is over~$4$~ms. The reason is that the need for backtracking increases with the physical link delay, and VMs become more likely to be scaled to their maximum capacity, which results in a higher cost. As one might expect, the \ratio for \bestfit decreases when physical link delay $\ge 4$~ms because it does not serve requests of higher cost service $s_1$. Recall that the cost of a service depends on the amount of required CPU on VMs, and therefore services with lower target delays incur more costs to serve one unit of traffic.

\begin{figure}[t]
	\centering
	\includegraphics[width=1.0\linewidth]{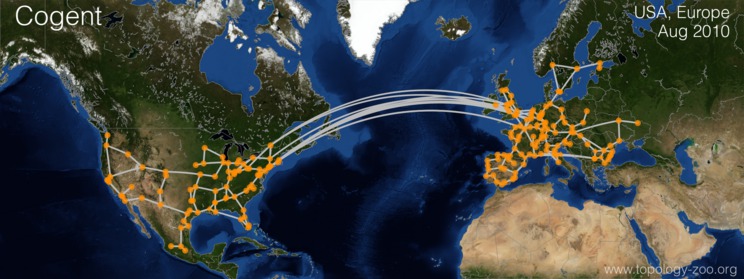}
	\caption{Cogent Network Topology.}
	\label{fig:cogent}
\end{figure}

{\bf Large-scale scenario.}
We consider the real-world inter-datacenter network \textit{Cogent}, a tier~$1$ Internet service provider (\Fig{cogent}). This network topology contains $197$ access nodes with $245$ physical links and $32$ datacenters. We set the cost of links connecting the datacenters to $0.02$~\euro/GB. The delay of logical links connecting the datacenters is set to be proportional to their geographical lengths, while the links inside each datacenter are assumed to be ideal having no capacity limit, latency, and cost. 
We assume each datacenter hosts $42$ VMs, each of which is connected to some edge switches. We categorize VMs within each datacenter in \textit{small}, \textit{medium}, and \textit{large} types according to their capacity and cost, as described in \Tab{vm-types}. We assume VMs need one minute to setup before being \textit{active}.

We consider the four different services described in~\Tab{services}, each of which is a representative of a category of real 5G applications. In this scenario, we assume that the VNFFG of each service is a chain of five VNFs. We further assume that the computational complexity, \ie, $\omega$ and maximum number of instances, \ie, $N(s, q)$ for two VNFs of service $s_4$ are three, while other VNFs have $\omega=1$ and $N(s, q)=1$. Similar to the previous scenario, we consider an equal number of requests for each service where requests arrive across time steps with the average inter-arrival time of three minutes and end after an average duration of two hours, and the total system lifespan is assumed to be one day. In this experiment, we set $H$ to $40$ minutes and $\tau$ to $20$ minutes.

\begin{figure}[t]
	\centering
	\subfloat[\label{fig:lambda-large-rev}]{{\includegraphics[width=0.5\linewidth]{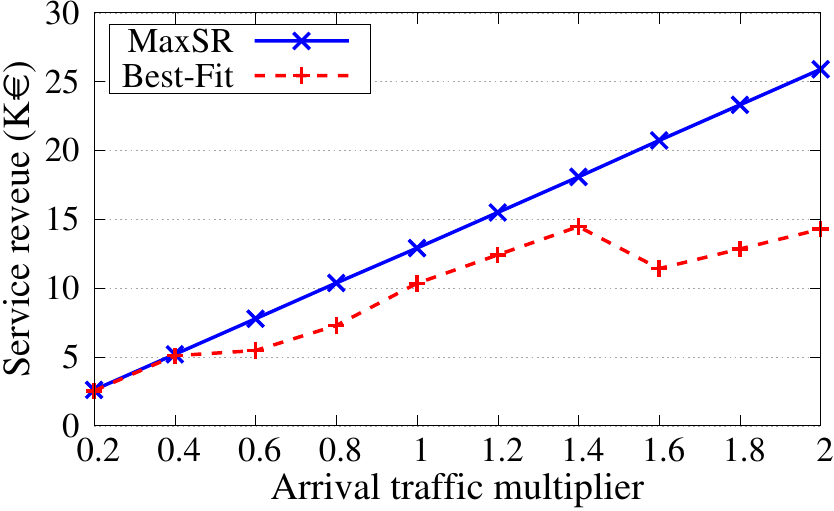} }}
	\subfloat[\label{fig:lambda-large-cost}]{{\includegraphics[width=0.5\linewidth]{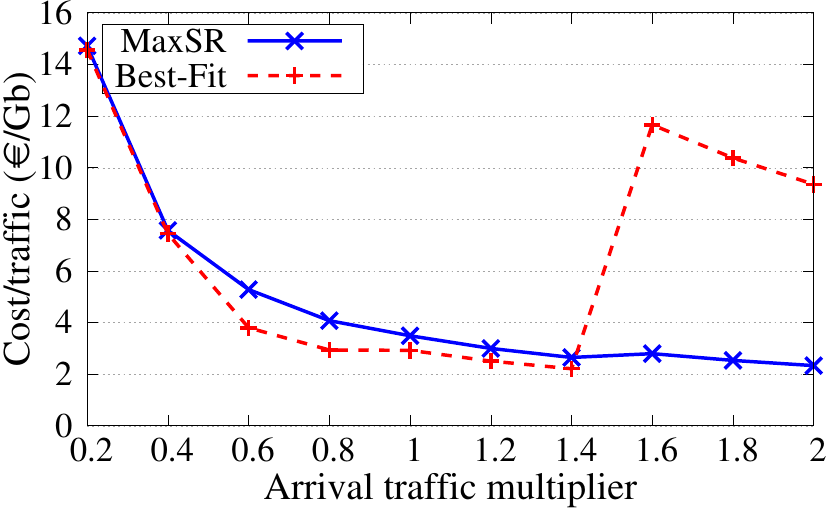} }}
	\caption{Large-scale scenario. Impact of service requests arrival traffic on absolute value of service revenue and \ratio. Physical link delay multiplier = $1$.}
	\label{fig:lambda-large}
\end{figure}

\begin{figure}[t]
	\centering
	\subfloat[Traffic multiplier = $1.0$\label{fig:lambda-large-service1}]{{\includegraphics[width=0.5\linewidth]{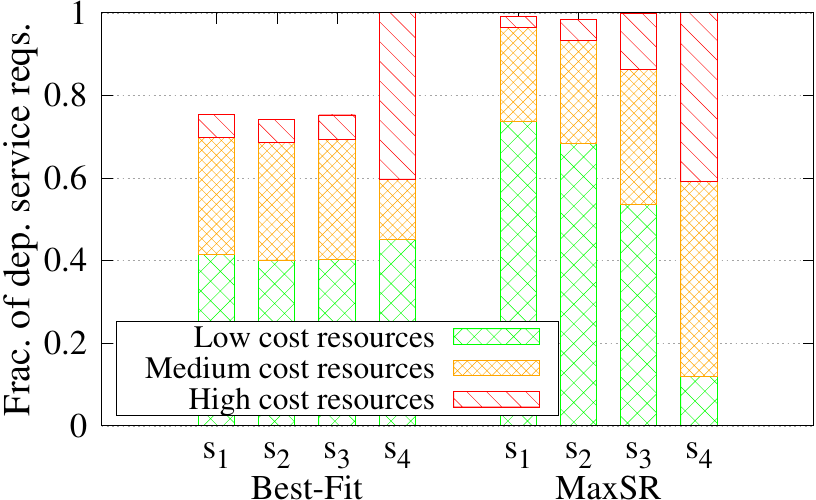} }}
	\subfloat[Traffic multiplier = $1.6$\label{fig:lambda-large-service2}]{{\includegraphics[width=0.5\linewidth]{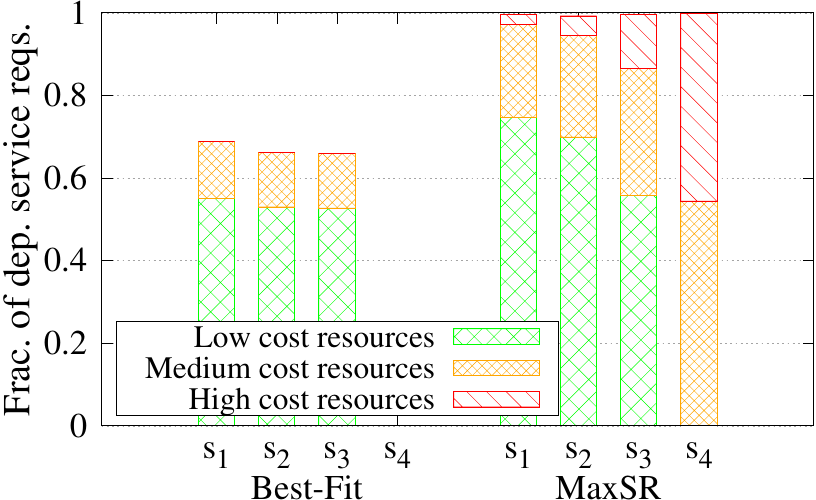} }}
	\caption{Fraction of deployed service requests for each service and algorithm. Physical link delay multiplier = $1$.}
	\label{fig:lambda-large-service}
\end{figure}

{\bf Impact of Physical Link latency and Arrival Traffic.}
As explained above, the optimum values cannot be obtained for this scenario in a reasonable time and therefore we rely on results for \heu and \bestfit. 
\Fig{lambda-large-rev} shows the effect of arrival traffic on the service revenue, while \Fig{lambda-large-service} shows the fraction of requests of each service that can be successfully deployed. We observe that service revenue for \heu changes almost proportionally with the traffic because increasing the traffic almost does not impact the fraction of served requests by this algorithm. \bestfit serves a lower fraction of service requests, and therefore achieves lower revenue. Besides, \bestfit shows a drop-off in service revenue when the arrival traffic multiplier is $1.6$: as confirmed by \Fig{lambda-large-service2}, this is because \bestfit does not serve requests of high traffic service $s_4$ when the traffic multiplier is over~$1.6$, due to its lack of support for multiple VNF instances.

\begin{figure}[t]
	\centering
	\subfloat[Traffic multiplier = $1.0$\label{fig:lambda-large-traffic1}]{{\includegraphics[width=0.5\linewidth]{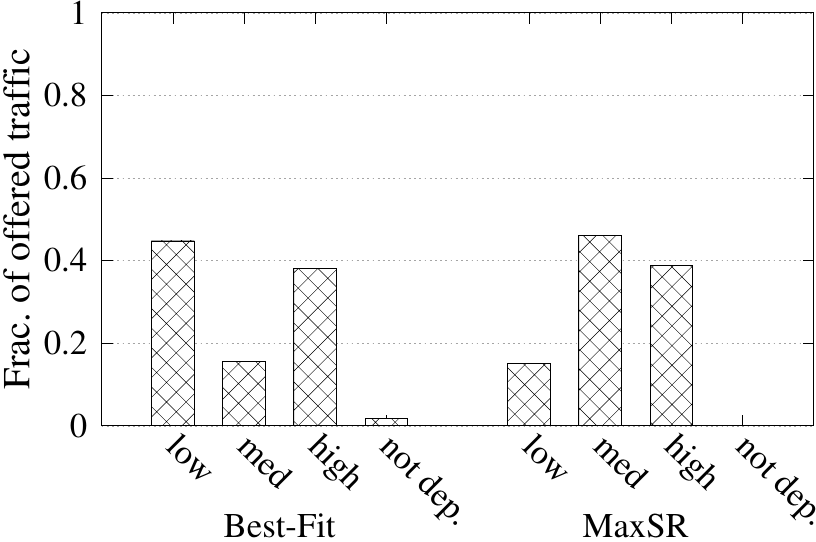} }}
	\subfloat[Traffic multiplier = $1.6$\label{fig:lambda-large-traffic2}]{{\includegraphics[width=0.5\linewidth]{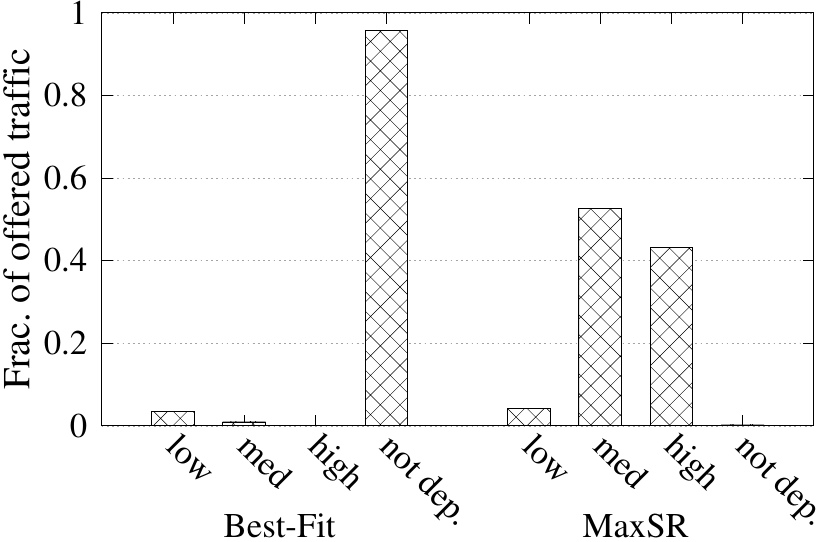} }}
	\caption{Fraction of offered traffic deployed on each resource type for each algorithm. \textit{low}, \textit{med}, \textit{high} and \textit{not dep.} mean the fraction of offered traffic served on low cost resources, medium cost resources, high cost resources, and which is not served, respectively. Physical link delay multiplier = $1$. }
	\label{fig:lambda-large-traffic}
\end{figure}

\Fig{lambda-large-cost} shows the impact of arrival traffic on the \ratio. \bestfit has lower \ratio when arrival traffic multiplier is less than~$1.4$, since \heu must use resources with higher cost and higher computational capabilities to serve more service requests; in other words, \bestfit serves less traffic but that traffic is served cheaply. Similar to service revenue the values of \ratio for \bestfit have a significant rise when the arrival traffic multiplier is~$1.6$, as confirmed by \Fig{lambda-large-traffic}, when the traffic multiplier increases from $1.0$ to $1.6$, \bestfit is no longer able to serve a significant fraction of the total traffic. As shown by \Fig{lambda-large-service}, the traffic \bestfit is unable to serve mainly belongs to the low cost service $s_4$, which results in a higher cost for served traffic.

\begin{figure}[t]
	\centering
	\subfloat[\label{fig:delay-large-rev}]{{\includegraphics[width=0.5\linewidth]{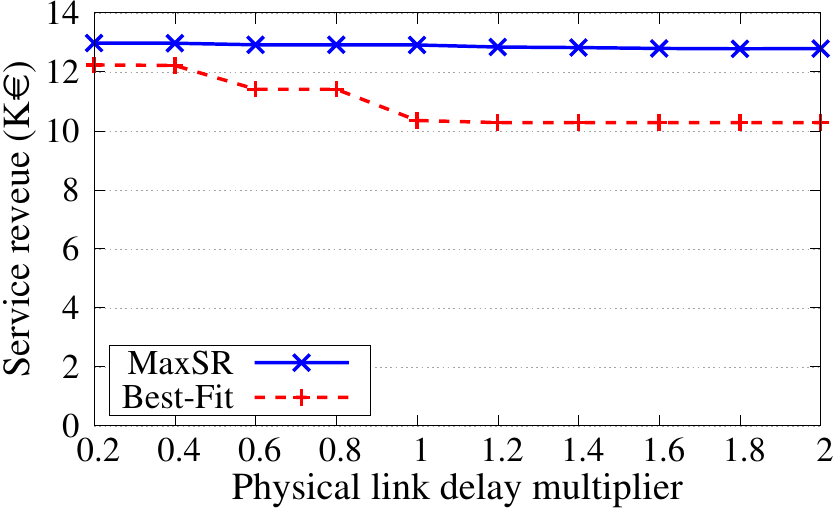} }}
	\subfloat[\label{fig:delay-large-cost}]{{\includegraphics[width=0.5\linewidth]{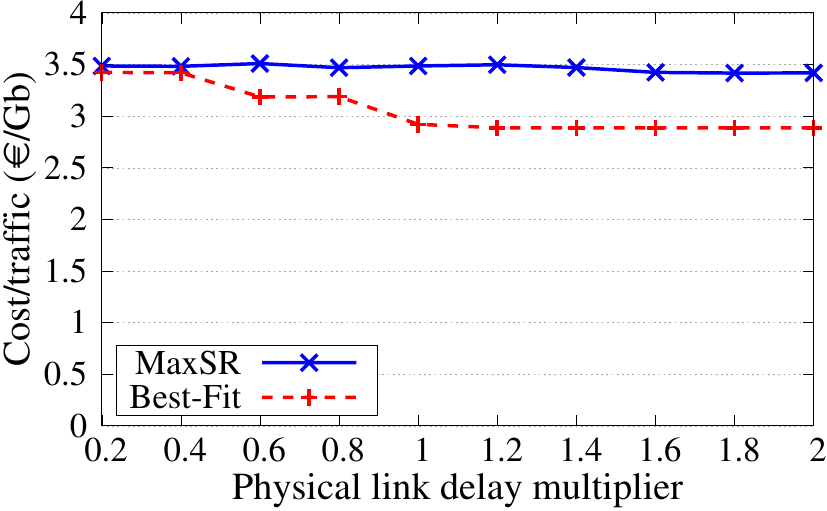} }}
	\caption{Large-scale scenario. Impact of physical link delay on absolute value of service revenue and \ratio. Arrival traffic multiplier = 1.}
	\label{fig:delay-large}
\end{figure}

\Fig{delay-large-rev} shows the impact of physical link latency on the service revenue. Similar to the small-scale scenario, \heu outperforms \bestfit especially for higher values of physical link delays, because the latter cannot serve the requests for services with low target delay (hence, higher revenue). As these service types have higher cost, they will also cause the \ratio for \bestfit to be lower than \heu as shown in \Fig{delay-large-cost}.

\begin{figure}[t]
	\centering
	\subfloat[Physical link delay multiplier = $0.6$\label{fig:delay-large-service1}]{{\includegraphics[width=0.5\linewidth]{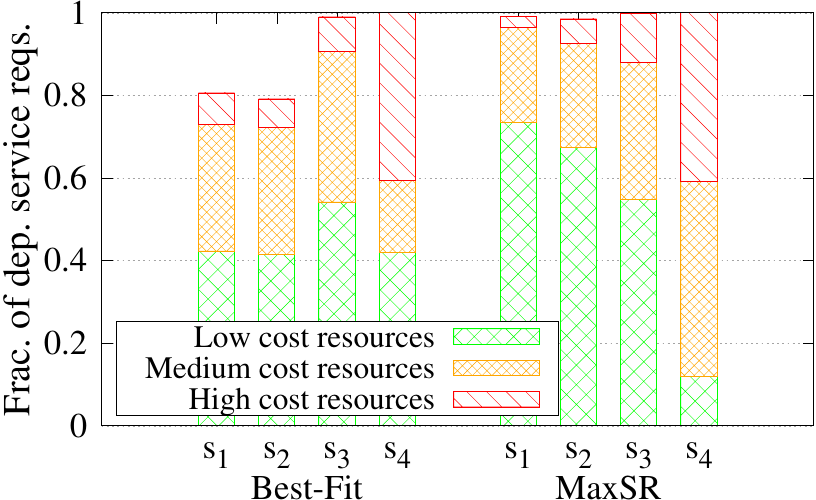} }}
	\subfloat[Physical link delay multiplier = $1.2$\label{fig:delay-large-service2}]{{\includegraphics[width=0.5\linewidth]{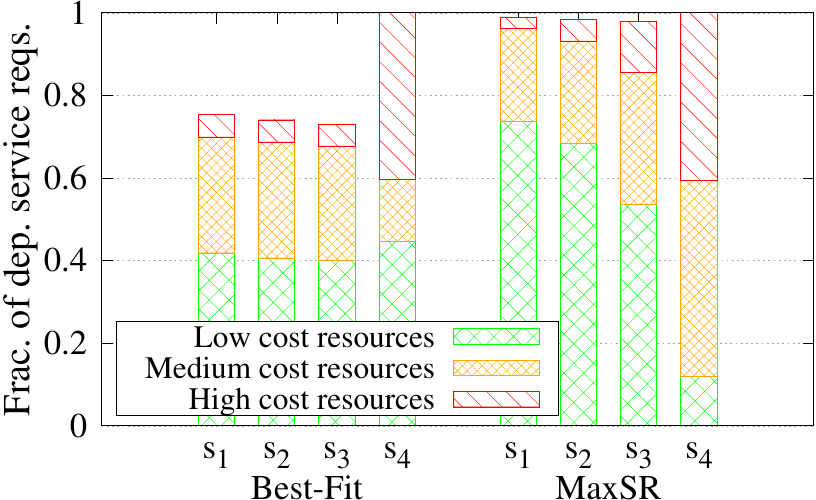} }}
	\caption{Fraction of deployed service requests for each service and algorithm. Arrival traffic multiplier = $1$.}
	\label{fig:delay-large-service}
\end{figure}

{\bf Running Time.} 
We run our experiments using a server with $40$-core Intel Xeon E5-2690 v2 3.00GHz CPU and $64$~GB of memory. To compare the running time of different algorithms, we consider the case where the arrival traffic and physical link delay multipliers are equal to one. For each scenario, we run the algorithm $50$ times and report the average running time in \Tab{execution-time}. \heu and \bestfit are substantially faster than brute-force in the small-scale scenario. The prohibitively long running time for brute-force highlights its poor scalability, and makes it inapplicable for the large-scale scenario in practice. The results for the large-scale scenario show that although \heu has higher running time compared to \bestfit due to backtracking, both of them are scalable and adequately fast for large-scale networks.

\begin{table}[t]
	\caption{Running time (in seconds)}	
	\label{table:execution-time}	
	\centering	
	\begin{tabular}{cccc}	
		\toprule	
		Scenario & Brute-force & \heu & \bestfit \\ 	
		\midrule	
		Small-scale & $399$ & $0.2$ & $0.14$ \\ 	
		Large-scale & - & $21$ & $2$ \\	
		\bottomrule	
	\end{tabular}	
\end{table}

\section{Conclusion\label{sec:conclusion}}

We proposed a dynamic service deployment strategy in 5G networks, accounting for real-world aspects such as VM setup times, and jointly making all the required decisions. We first formulated the problem of joint requests admission, VM activation, VNF placement, resource allocation, and traffic routing as a MILP based on the complete knowledge of requests arrival and departure times. We took the MNO profit as the main objective to be optimized over the entire system lifespan, leveraging a queueing model to ensure all requests adhere to their latency targets. Our model also accounted for the key features of 5G services such as complex VNF graphs and arbitrary input traffic. 

Due to the problem complexity, we further proposed a heuristic, \heu, which has polynomial complexity and attains near-optimal solutions, while only needing the knowledge/prediction of the upcoming service requests in a short time horizon. The algorithm works in a sliding-horizon fashion, rearranging the current-served requests across existing VMs to reduce the deployment costs, and admitting the new ones as they arrive at the system. Furthermore, the parameters of \heu allow for different tradeoffs between solution optimality and running time. We demonstrated the effectiveness and efficiency of our approach through a numerical evaluation including different network scenarios. 


\section*{Acknowledgments}
This work was supported by the EU 5GROWTH project (Grant No. 856709).

\bibliographystyle{IEEEtran}
\bibliography{IEEEabrv,refrences}

\end{document}